\documentclass[a4paper,fleqn,usenatbib]{mnras}

\usepackage{newtxtext,newtxmath}

\usepackage{graphicx}	
\usepackage{amsmath}	
\usepackage{amssymb}	

\usepackage[T1]{fontenc}
\usepackage{ae,aecompl}
\usepackage{pdflscape}
\usepackage{multicol}
\usepackage{bm}
\usepackage{subfigure}
\usepackage{pstricks}

\title[Degradation of photometric redshifts with non-representative training sets]{Degradation analysis in the estimation of photometric redshifts from non-representative training sets}

\author[J. D. Rivera et al.]{
J. D. Rivera,$^{1}$\thanks{E-mail: \href{mailto:josedavidr@ift.unesp.br}{josedavidr@ift.unesp.br} (JDR)}
B. Moraes,$^{2}$
A. I. Merson,$^{3,4}$
S. Jouvel,$^{2}$
F. B. Abdalla$^{2,5}$\thanks{E-mail: \href{mailto:fba@star.ucl.ac.uk}{fba@star.ucl.ac.uk} (FBA)}
\newauthor
and M.C.B Abdalla$^{1}$ 
\\\\
$^{1}$Instituto de F\'isica Te\'orica, Universidade Estadual Paulista, R. Dr. Bento Teobaldo Ferraz 271, S\~ao Paulo 01140-070, Brazil\\
$^{2}$Department of Physics and Astronomy, University College London, Gower Street, London WC1E 6BT, UK\\
$^3$Jet Propulsion Laboratory, California Institute of Technology, 4800 Oak Grove Drive, Pasadena, CA 91109, USA\\
$^4$IPAC, Mail Code 314-6, California Institute of Technology, 1200 East California Boulevard, Pasadena, CA 91125, USA\\
$^5$Department of Physics and Electronics, Rhodes University, PO Box 94, Grahamstown, 6140, South Africa
}

\date{Accepted XXX. Received YYY; in original form ZZZ}

\pubyear{2017}

\begin{document}
\label{firstpage}
\pagerange{\pageref{firstpage}--\pageref{lastpage}}
\maketitle

\begin{abstract}
We perform an analysis of photometric redshifts estimated by using a non-representative training sets in magnitude space. 
We use the \texttt{ANNz2} and \texttt{GPz} algorithms to estimate the photometric redshift both in simulations as well as in real data from the Sloan Digital Sky Survey (DR12). 
We show that for the representative case, the results obtained by using both  algorithms have the same quality, either using magnitudes or colours as input.
In order to reduce the errors when estimating the redshifts with a non-representative training set, we perform the training in colour space. 
We estimate the quality of our results by using a mock catalogue which is split samples cuts in the $r$-band between $19.4< r< 20.8$. 
We obtain slightly better results with \texttt{GPz} on single point z-phot estimates in the complete training set case, however the photometric redshifts estimated with \texttt{ANNz2} algorithm allows us to obtain mildly better results in deeper $r$-band cuts when estimating the full redshift distribution of the sample in the incomplete training set case. 
By using a cumulative distribution function and a Monte-Carlo process, we manage to define a photometric estimator which fits well the spectroscopic distribution of galaxies in the mock testing set, but with a larger scatter.
To complete this work, we perform an analysis of the impact on the detection of clusters via density of galaxies in a field by using the photometric redshifts obtained with a non-representative training set.
\end{abstract}




\section{Introduction}

The cosmological redshift of a galaxy is arguably one of the most important directly observable properties and provides a measure of the recessional velocity of that galaxy, relative to an observer, which arises due to the expansion of the Universe.
In General Relativity, knowledge of the redshift of an object allows one to connect the spatial and time-dependent components of the space-time metric. 
A cosmological model provides us with a prediction of how to accurately translate between the redshift of an object and the physical distance to that object.  
As such, a precise measurement of this relation allows us to place tight constraints on cosmological parameters \citep{riess,perlmutter}, and therefore on our fundamental understanding of cosmology. 
This is a major goal of future cosmological missions that aim to make high precision measurements of various \emph{cosmological probes}; including measurements of Baryon Acoustic Oscillations (BAO, \citealt{HuDodelson,eisenstein,percival,blake,anderson}), the weak lensing of galaxies \citep{massey,bartelmann,kilbinger}, and the number counts of galaxy clusters \citep{battye,mantz,rozo,allen,mana}.

The redshift of a galaxy can be measured in two ways: either \emph{spectroscopically} or \emph{photometrically}. Spectroscopic determination of redshift involves measuring the Doppler shift of known features in the spectrum of a galaxy, typically absorption or emission lines. 
Photometric determination of redshift is based upon the assumption that the colours of a population of galaxies of the same type and redshift (i.e. with very similar spectra) will be clustered in a particular region of the colour space. 
One can therefore estimate the photometric redshift of a galaxy by using multi-band photometry to compare the broad-band colours of that galaxy with the colours of set of galaxies for which redshifts are already known \citep[see][]{benitez, collister, ilbert, almosallam, sadeh}. Since the measurement of the spectrum of a galaxy is much more costly, photometric redshifts provide a cheaper and rapid alternative though are less accurate than spectroscopic redshift determination. Therefore, photometric redshifts are a viable and efficient option to be used in cosmological surveys that plan to observe several billion galaxies, including the Dark Energy Survey (DES)\footnote{\label{DES_survey}\url{http://www.darkenergysurvey.org}}, the Large Synoptic Survey Telescope (LSST)\footnote{\label{LSST_survey}\url{http://www.lsst.org}}, the Euclid\footnote{\label{Euclid_survey}\url{http://sci.esa.int/euclid/}} and the Wide Field Infrared Survey Telescope (WFIRST)\footnote{\label{WFIRST_survey}\url{https://wfirst.gsfc.nasa.gov}}. 
Note that the Euclid and WFIRST missions will additionally measure spectroscopic redshifts for a sub-set of galaxies. 
The major challenge that these surveys face is the problem that photometric redshifts are much less precise than spectroscopic redshifts and will need considerable calibration. 

We can split the photometric techniques into two approaches: machine learning and template fitting. 
Machine learning involves using machine learning methods (MLMs) to establish the relationship between the photometric observables (e.g. colours or magnitudes) and the redshift of a galaxy. 
This is usually done by training these methods on a dataset of galaxies with known redshifts. 
Among these methods we have the artificial neural networks \citep[ANNs,][]{firth, vanzella}, which include the codes \texttt{ANNz} \citep{collister} and \texttt{ANNz2} \citep{sadeh}; nearest-neighbour techniques \citep{ball}; random forest techniques, including \texttt{TPZ} \citep{carrasco}; and Gaussian processes \citep[GPs,][]{way, bonfield, almosallam1} such as the \texttt{GPz} \citep{almosallam} code. 
The effectiveness of these methods depends on whether the training set is a representative sample of the photometric dataset.
Moreover, the MLMs are only reliable over the redshift range of the training data set that is used.
Therefore, in principle, those methods cannot be employed to estimate high redshifts for which no spectroscopic data is available.
Template methods are based on fitting empirical or synthetic galaxy spectra with the photometric information available (i.e., colours or magnitudes).
Specifically, they use the broad-band photometry to estimate an approximate galaxy spectral energy distribution (SED), which they then fit against a library of SEDs with known redshifts.
Those methods require astrophysical effects, e.g. the dust extinction in the observed galaxy or in our galaxy to be corrected for.
A non exhaustive list of codes known for template fitting methods are \texttt{HYPERZ} \citep{bolzonella}, \texttt{ZEBRA} \citep{feldmann}, \texttt{EAZY} \citep{brammer} and \texttt{LE PHARE} \citep{arnouts, ilbert}. 
Both techniques to estimate photometric redshifts have advantages and limitations depending on the spectroscopic data available and the photometric data set to being evaluated.
\cite{abdalla2008,hildebrandt,abdalla} and \cite{sanchez} have compared different photometric redshift techniques and their efficiency in ground and space data.

The lack of the large sets of spectroscopic data to train and validate the photometric surveys of galaxies is a critical problem nowadays.
\cite{beck} performed a quantitative analysis of the problem described above by using machine learning and template fitting approaches in real data.
They performed several tests in order to assess the reliable of the used methods to estimate photometric redshift in the cases with representative and unrepresentative spectroscopic data set used for train the validation/testing set as well as the effects because of the photometric error.
\cite{cavuoti} tackled the accuracy problem for estimated photometric redshifts by using SED fitting and machine learning methods on The Kilo Degree Survey (KiDS) photometric galaxy data \citep{de_Jong}. 
They showed that in the representative regions of the input parameters for the training set, the empirical method provide better results. 
Nonetheless, the theoretical method provide information about the galaxy spectral type. Concluding in this way that the hybrid technique combining template fitting and MLM might improve the z-phot prediction accuracy.
Unlike to \cite{beck} and \cite{cavuoti}, our main aim is to set the fainter flux limit for galaxies in which we achieve to estimate reliable photometric redshifts on a realistic mock catalogue of the SDSS survey, when the training data set is non-representative of the testing data set in the magnitude space.
In particular we examine the degradation in the redshifts obtained when the testing data set extends to limiting magnitudes significantly fainter than the training data set. 
We first examine the degradation using a mock catalogue, before performing a comparable analysis using real data. Here we use the \texttt{ANNz2} and \texttt{GPz} algorithms, which belong to group of machine learning techniques. 
We also compare the single value estimators obtained from the calculated PDF photometric redshift, the classical mean value and another based on a Monte-Carlo sampling from the cumulative function. 
We show that the latter one is the best to represent the estimated redshift distribution in agreement with the true redshift on the mock catalogue.
On the other hand, we perform an analysis on the impact
in the detection of galaxy clusters via density methods, such as Voronoi Tessellation or kernel density \citep{gal,lopes,soares}, by using photometric redshifts estimated by these non-representatives training sets.

We organise the paper as follows: in section~\ref{catalogues}, we present the mock catalogue and the observational datasets (including the training set) which are used in this analysis. In section~\ref{methods_zphot} we describe the \texttt{ANNz2} and \texttt{GPz} algorithms used in this work and introduce the metrics used to assess the quality of the derived photometric redshifts. Both of these algorithms output for each galaxy a single redshift estimate as well as a redshift probability distribution function (PDF). As such we also introduce two estimators to additionally compute the photometric redshifts using the full PDF information. In section~\ref{results} we compare, for both the mock catalogue and observed datasets, the quality of the derived photometric redshifts obtained using the \texttt{ANNz2} and \texttt{GPz} algorithms and examine the impact of building our training set using either magnitude-space or colour-space selection criteria. We then apply sequentially deeper $r$-band magnitude cuts to the mock catalogue in order to analyse the degradation in the quality and completeness of the derived photometric redshifts when the testing set extends to $r$-band magnitudes significantly deeper than the training set. In section~\ref{implication_galaxy_clusters} we discuss the impactions that this has on the detection of galaxy clusters. Finally, in section~\ref{conclusions} we summarise our conclusions.

\section{Data}
\label{catalogues}

In order to assess the robustness of our results, we use consider first a simulated data set, before analysing a real, observed data set. Simulated galaxies are taken from a lightcone mock catalogue constructed from a galaxy formation model. The advantage of first using a mock data set is that we can measure the precision and accuracy of the estimated photometric redshifts for a population of galaxies for which the true redshifts are already known. We then apply our methods to observed galaxy data sets extracted from the Sloan Digital Sky Survey Data Release 12 (SDSS DR12). In this instance a training data set with known spectroscopic redshifts is taken from the Galaxy And Mass Assembly (GAMA) survey. For consistency, we construct the photometric SDSS testing set by applying the GAMA selection criteria to the SDSS DR12 galaxies. 
Here we describe the mock and real data sets in more detail.

\subsection{Mock galaxy catalogue}
\label{mock_catalogue}

The mock catalogue\footnote{\label{catalogue_SDSS_500} We use the SDSS\_500\_photoz catalogue available from \url{http://astro.dur.ac.uk/~d40qra/lightcones/SDSS/}.} used in this work was constructed using the lightcone construction method presented in \cite{Merson2012}. 
In brief, this method involves populating the dark matter halo merger trees extracted from a cosmological N-body simulation with galaxies generated from a \emph{semi-analytical} galaxy formation model. 
In this case, the merger trees were taken from the Millennium Simulation \citep{Springel2005} and populated using the \cite{Lagos2012} version of the \texttt{GALFORM} model, which was originally developed by \cite{Cole2000}. A lightcone catalogue is then constructed by interpolating the galaxy positions between the simulation redshift snapshots to determine when each galaxy crosses the past lightcone of the observer. For further details we refer the reader to \cite{Merson2012}. The cosmology used in the Millennium Simulation is a $\Lambda$ cold dark matter (CDM) model ($\Omega_m$, $\Omega_{\Lambda}$, $\Omega_b$, $h$ = 0.25, 0.75, 0.045, 0.73), with parameters consistent with the first year results from the Wilkinson Microwave Anisotropy Probe \citep{Spergel03}.

The lightcone catalogue spans the redshift range $z=0.0$ to $z=3.0$ and has a sky footprint of approximately $500\,\mathrm{deg}^2$, centred on position (RA, DEC) $\simeq$ (303.29 deg, -14.48 deg). An SDSS $r$-band selection ($r\leqslant 24$) was applied to the lightcone, yielding a total of 15 823 757 galaxies. The $(u,g,r,i,z)$ magnitudes of galaxies reported in the lightcone are AB apparent magnitudes. 
For each photometric band, $X$, the magnitudes are perturbed to introduce photometric noise by randomly sampling from a Gaussian with a mean, $m_X$, equal the AB apparent magnitude of the galaxy in that band, and with a standard deviation, $\sigma_X(m_X)$, which is defined following the approach described in \cite{Jouvel09} as,
\begin{equation}
\sigma_X=    
\begin{cases}
10^{0.4(\gamma_o+1)\left (m_X-m^{\star}_X\right )}, & \text{if}\ m_X<m^{\star}_X, \\
\frac{\sigma^{\star}}{2.72}\exp\left (10^{\gamma_s\left (m_X-m^{\star}_X\right )}\right ), & \text{otherwise},
\end{cases}
\end{equation}
where $m^{\star}_X$ is a characteristic magnitude, $\sigma^{\star}$ is a normalisation coefficient and $\gamma_o$ and $\gamma_s$ are power-law slopes. The values adopted for these parameters are shown in Table~\ref{err_phot}. The power-law used in the case $m_X<m^{\star}_X$ corresponds to brighter fluxes, dominated by object noise, whilst the exponential law in the case $m_X \geqslant m^{\star}_X$ corresponds to fainter fluxes dominated by sky background noise.
For further details see \cite{Jouvel09}.
In order to obtain a sample similar to our GAMA/SDSS data set, we apply a further $i$-band magnitude cut $i<21$, which leaves a total of 1 876 505 galaxies, with a mean redshift of $z_{\mathrm{mean}}\sim 0.35$.

\begin{table}
\centering
\caption{Values for the characteristic magnitude ($m^{\star}_X$), the normalisation coefficient ($\sigma^{\star}$), the bright magnitude slope ($\gamma_o$) and faint magnitude slope ($\gamma_s$) used to compute photometric noise in each photometric band ($X$) in the SDSS mock data. See text in section~\ref{mock_catalogue} for details. The magnitude limit for the $u$-band is from \citet{zou} and the magnitude limits for the $g$-band, $r$-band, $i$-band and $z$-band are from \citet{raichoor}.}
\label{err_phot}
\begin{tabular}{c c c c c}
\hline \hline
$X$ & $m^{\star}_X$ & $\sigma^{\star}$ & $\gamma_o$ & $\gamma_s$ \\ \hline 
 $u$ & $22.03$ & $0.2$ & $-0.1$ & $0.25$ \\ 
 $g$ & $23.10$ & $0.2$ & $-0.1$ & $0.25$ \\ 
 $r$ & $22.70$ & $0.2$ & $-0.1$ & $0.25$ \\ 
 $i$ & $22.20$ & $0.2$ & $-0.1$ & $0.25$ \\  
 $z$ & $20.70$ & $0.2$ & $-0.1$ & $0.25$ \\ \hline 
\end{tabular}
\end{table}

\subsection{GAMA survey}

The Galaxy and Mass Assembly (GAMA) survey\footnote{\label{GAMA_survey}\url{http://www.gama-survey.org/}} consists of optical spectroscopy data for the low-redshift galaxy popu\-lation. 
The survey was designed to investigate galaxy formation and evolution processes, occurring within the galaxy distribution on scales of 1 kpc to 1 Mpc \citep{driver2009, baldry}.
Observations were performed with the AAOmega spectrograph on the Anglo-Australian Telescope (AAT), covering a sky area of $\sim 286\, \mathrm{deg}^2$ split into five survey regions on the sky, with a total of 238 000 objects.
The regions observed were split into three equatorial regions (G09, G12, G15) and two southern sky regions (G02 and G23) \citep{driver2011, liske}. 

The survey consisted of two phases, each with slightly different target selection criteria. GAMA I refers to data collected during the first three years, while GAMA II refers to the full survey, including all of GAMA I. The first phase extended over the three equatorial regions down to (extinction-corrected) Petrosian magnitude of $r_{\mathrm{petro}}<19.4$ in G09 and G15, and $r_{\mathrm{petro}}<19.8$ in G12. 
Magnitude cuts and target selection were based on photometry from SDSS and additional infrared bands from the UKIRT (United Kingdom InfraRed Telescope) Infrared Deep Sky Survey (UKIDSS), which were introduced to help improve star-galaxy separation.
In the second phase, the three existing equatorial survey regions were enlarged and the two southern regions, G02 and G23, were added. The $r$-band Petrosian magnitude limit was pushed to $r_{\mathrm{petro}}<19.8$ in all survey regions.

Here we use the public Data Release 2 (DR2), which includes the galaxies from GAMA I of survey region G15 ($r_{\mathrm{petro}}<19.4$) and a subset of G09 and G12 survey regions ($r_{\mathrm{petro}}<19.0$) with a total area of $\sim 144\, \mathrm{deg}^2$ for a total of 70 726 targets with secure redshifts download from the GAMA database. For more details, see \cite{baldry} and \cite{liske}. To match to the selection criteria of our photometric sample, we then use the SDSS DR12 CasJobs server\footnote{\label{SDDS_CasJobs}\url{https://skyserver.sdss.org/CasJobs/}} to cross-match the GAMA data to a clean (i.e. \texttt{CLEAN=1}) sub-sample of SDSS DR12 galaxies with additional ``GAMA-like" cuts. Our final spectroscopic sample contains 63 226 objects with $r_{\mathrm{petro}}<19.4$.

\subsection{SDSS DR12 sample}
\label{sdss_dr12}

Our photometric data set is obtained from a parent sample downloaded from the Sloan Digital Sky Survey Data Release 12 (SDSS DR12) database. Since we consider the GAMA survey as the spectroscopic training sample, the choice of photometric data is performed by using the GAMA target selection cuts in the SDSS DR12 according to \cite{christodoulou}. Here we consider two cases for our analysis.

In the first case we use the magnitude and colour cuts, such that the training set is a fully representative in the magnitude space. We shall refer to this sample as GAMA MAIN.
In the second case we relax the magnitude limit but keep the sample fully representative in colour space. For this sample, the training is performed with 4 colours instead of 5 magnitudes. We shall refer to this second sample as GAMA DEEP.

The rationale for the choice of 4 colours is as follows: training with $ugriz$ magnitudes occurs in a 5-dimensional data space. This would be equivalent to a 5-dimensional training with 4 colours and 1 magnitude - it is, in fact, a linear transformation from the space of 5 magnitudes. By restricting ourselves to an arbitrary set of 4 colours, we are excluding the dimension where our sample is not representative, with the expectation that redshifts are mostly correlated with colours. This assumption is not completely accurate and will depend on how far we push the magnitude limit fainter than the spectroscopic sample. In short, this new training criterion is chosen as to ignore the non-representativeness of the training in magnitude space. In Appendix~\ref{SQL_query} we show the SQL query used to obtain GAMA MAIN and GAMA DEEP samples from the SDSS DR12 database.
See Moraes {\it et al.} 2017 (in prep.) for more details about these samples.

\section{Estimating photometric redshifts}
\label{methods_zphot}

In order to estimate the photometric redshifts for galaxies in the GAMA and SDSS surveys and as well as the mock catalogues, we use the \texttt{ANNz2} \citep{sadeh} and \texttt{GPz} \citep{almosallam} public photometric redshift algorithms. 
These codes apply a set of machine learning methods, using a set of training redshifts to estimate the value of redshift for galaxies without spectroscopic information from their photometry. 
We briefly describe the \texttt{ANNz2} and \texttt{GPz} codes. 

\subsection{ANNz2}
\label{ANNz2_desc}

\texttt{ANNz2}\footnote{\label{ANNz2_code}\url{https://github.com/IftachSadeh/ANNZ}} \citep{sadeh} is a updated version of the original \texttt{ANNz} package developed by \cite{collister}, which used artificial neural networks (ANNs) to estimate the photometric redshifts of galaxies. Given a training set of galaxies, \texttt{ANNz2} combines different machine learning techniques (i.e., artificial neural networks, boosted decision/regression trees, among others) to compute a photometric redshift probability distribution function (PDF) for each galaxy in the testing set. 
The machine learning methods (MLMs) employed are implemented in the TMVA package \citep{hoecker}. 

Like all MLMs, the \texttt{ANNz2} code requires training and validation samples from a spectroscopic redshift survey.
During each step of the training, the validation sample is used to estimate the convergence of the solution. 
Once the mapping is established, an independent testing set (i.e., an independent subsample from the spectroscopic redshift survey with photometric information) is used to evaluate the performance of the trained MLM. 
The methods implemented in this code allow us to optimise the photometric redshift reconstruction, and to estimate their associated uncertainties, which helps mitigate possible problems of non-representativeness.
To correct for inaccuracies due to unrepresentative training sets, the \texttt{ANNz2} algorithm can use training weights. 
This method aims to match the distribution of the inputs from the training sample with the testing data following the approach presented in \cite{lima}.
If the training data set is incomplete (i.e., there are some regions of the input phase-space where the evaluated sample has no corresponding objects for training), this code provides a quality flag, which indicates when unrepresented data are being evaluated.

In order to estimate the photometric redshift PDFs for galaxies, the \texttt{ANNz2} algorithm follows an approach called randomised regression, which ranks the different solutions according to their performance based upon the values of va\-rious metrics (i.e., bias, scatter, level of outliers). The entire set of solutions is used to construct the photometric redshift PDF. Initially, each solution is folded with a distribution of uncertainty values computed via the \emph{K-nearest neighbours} (KNN) method \citep[see][]{oyaizu}. 
The values that we specified for the input parameters to \texttt{ANNz2} are provided in Table~\ref{param_ANNz2_GPz}.

\subsection{GPz}
\label{GPz_desc}

\texttt{GPz}\footnote{\label{GPz_code}\url{https://github.com/OxfordML/GPz/blob/master/python/demo_photoz.py}} \citep{almosallam} is a machine learning approach which uses sparse Gaussian processes (GPs) to estimate a photometric redshift and its variance. 
The GPs are probabilistic models for regression. 
The assumption underlying the GP approach is that there exists a function, $f$, to map a set of target inputs $\mathbf{X}$ (i.e. the galaxy photometry) onto a set of target outputs $\mathbf{Y}$ (i.e. the galaxy redshifts), such that $\mathbf{Y}=f(\mathbf{X})+\epsilon$, where $\epsilon$ is an additive noise that is assumed to be Gaussian. GPs are probabilistic models for regression with which we can construct a probability distribution for the possible forms of the function $f$.
The computational cost for training these methods can be very high, towards impractical in certain instances. The problem is that the training depends on the cost required to invert an $n\times n$ covariance matrix for a training sample with $n$ components. 
Different authors have proposed several techniques in order to reduce this problem. For example, \cite{zhang} showed that in some cases the covariance matrix could have a Toeplitz structure, which would relieve the cost in the inversion. 
\cite{tsiligkaridis} simplify the computation of the inversion by decomposing the covariance matrix into a sum of Kronecker products. However, such techniques cannot always be applied.

Another approach to relieve the computational cost is to reduce the size of the covariance matrix by using sparse approximations, such that the covariance matrix is obtained using a set of $m\ll n $ samples \citep[see][]{candela}. 
In \texttt{GPz} these sparse GPs are described using \emph{basis function models} (BFMs), with the function $f$ assumed to be a linear combination of these $m\ll n $ basis functions. 
In this method the variance is taken as an input-dependent function that is composed of two terms for different sources of uncertainty: the intrinsic uncertainty about the mean function due to data density and the uncertainty due to the intrinsic noise or the lack of precision in the training set.
Specifying the variance in this way is very useful to help identify regions of the input space where additional data is required, or where further precision is required, or both. With such information it is possible to develop strategies to increase the photometric accuracy, either by obtaining data in additional photometric bands or by improving the quality of input data in one particular photometric band or colour.

The \texttt{GPz} code can operate in different modes depending on the treatment of the covariance matrix for each basis function. These modes include,
\begin{itemize}
\item \texttt{GPVC}: The covariance matrix is different for each basis function (i.e. a GP with variable covariance).
\item \texttt{GPGC}: The covariance matrix is the same for all basis functions (i.e. a GP with global covariance).
\item \texttt{GPVD}: The covariance matrix is diagonal and different for each basis function (i.e. a GP with variable diagonal covariance).
\item \texttt{GPGD}: The covariance matrix is diagonal and the same for all basis functions (i.e. a GP with global diagonal covariance).
\item \texttt{GPVL}: The covariance matrix is given by $\Gamma_j=\textbf{I}\gamma_j$, where $\gamma_j$ is a scalar that is different for each basis function (i.e. a GP with variable length-scales).
\item \texttt{GPGL}: The covariance matrix is given by $\Gamma_j=\textbf{I}\gamma_j$, where $\gamma_j$ is a scalar that is the same for all basis functions (i.e. a GP with global length-scale). 
\end{itemize}
In Table~\ref{param_ANNz2_GPz} we show the parameter values used as input for the \texttt{GPz} code.

\begin{table*}
\centering
\caption{The used values for the parameters of the \texttt{ANNz2} (left) and \texttt{GPz} (right) codes.}
\label{param_ANNz2_GPz}
\begin{tabular}{c c c c c c c c}
\hline \hline
Code & Parameter & Definition & Value & Code & Parameter & Definition & Value \\ \hline 
 \texttt{ANNz2} & \texttt{nMLMs} & Number of MLMs & 100 & \texttt{GPz} & \texttt{method} & GP method & VC \\ 
  & \texttt{minValZ} & Min. value for redshift & 0.0 &  & \texttt{m} & Number of BFM & 25 \\ 
  & \texttt{maxValZ} & Max. value for redshift & 1.0 &  & \texttt{heteroscedastic} & Heteroscedastic noise & True  \\ 
  & \texttt{nErrKNN} & Near-neighbours for error & 90 &  & \texttt{csl\_method} & Cost-sensitive & Normal  \\  
 & \texttt{rndOptTypes} &  MLM types & ANN\_BDT &  & \texttt{maxIter} & Max. of iterations & 500   \\  
  & \texttt{nPDFbins} & Number of PDF bins & 200 &  & \texttt{maxAttempts} & Max. iterations to attempt & 50  \\  \hline 
\end{tabular}
\end{table*}

\subsection{Metrics}
\label{Metrics_work}

In order to assess the quality of photometric redshifts estimated in this work, we define the following set of commonly employed metrics to describe the bias and the scatter of the photometric redshifts relative to the spectroscopic redshifts, as well as the fraction of catastrophic outliers.

The bias measures the deviation of the estimated photometric redshift from the true value (i.e., the spectroscopic redshift). 
\begin{equation}
\label{bias_eqn}
\mathrm{Bias} = \left\langle\frac{z_{\mathrm{phot}}-z_{\mathrm{spec}}}{1+z_{\mathrm{spec}}}\right\rangle.
\end{equation}
The scatter between the true redshift and the photometric redshift is denoted as $\sigma$ and given by
\begin{equation}
\label{sigma_eqn}
\sigma = \left\langle\left(\frac{z_{\mathrm{phot}}-z_{\mathrm{spec}}}{1+z_{\mathrm{spec}}}\right)^2\right\rangle^{1/2}.
\end{equation}
We define $\sigma_{68}$, as
\begin{equation}
\label{sigma68_eqn}
\sigma_{68} = \max_{i\in\,U} \left\{\left|\frac{z_{\mathrm{phot}}^i-z_{\mathrm{spec}}^i}{1+z_{\mathrm{spec}}^i}\right|\right\},
\end{equation}
where $U$ is the set of the 68 percent of galaxies which have the smallest value of $|z_{\mathrm{phot}}-z_{\mathrm{spec}}|/(1+z_{\mathrm{spec}})$.
The catastrophic outlier rate, which we call $\mathrm{FR}_e$, is given by
\begin{equation}
\label{FRe_eqn}
\mathrm{FR}_e=\frac{100}{n}\left\{i:\left|\frac{z_{\mathrm{phot}}^i-z_{\mathrm{spec}}^i}{1+z_{\mathrm{spec}}^i}\right|<e\right\},
\end{equation}
where $n$ is the number of galaxies and $e$ is the outlier threshold.
This quantity is equal to the percentage of galaxies in the sample that are considered to have a good photometric redshift within a tolerance set using a chosen outlier threshold value. We choose $e=0.15$.

In order to compare the estimated photometric redshift distribution with the spectroscopic redshift distribution, we also define the chi-square measure $D_{\chi^2}$ as
\begin{equation}
\label{chi_dist}
D_{\chi^2}(P,Q)=\frac{1}{2}\sum_{i=1}^n\frac{\left[p(i)-q(i)\right]^2}{p(i)+q(i)},
\end{equation}
where $P(p_i)$ and $Q(q_i)$ are distribution functions and $p_i$ and $q_i$ are the variables in the different distributions. 
Note that if the two distributions are different, we obtain a high value for the chi-square measure.
Therefore, this metric allows us to quantify how similar is the distribution of the estimated photometric redshifts to that of the spectroscopic redshifts.

\subsection{Further photometric redshift estimators}
\label{photometric_estimators}

The \texttt{ANNz2} and the \texttt{GPz} codes provide for each galaxy both an individual redshift estimate as well as a full PDF. We describe here two estimators to extract a single redshift estimate based upon the full PDF information.

By integrating over the full PDF information we can estimate the mean photometric redshift, $z_{\mathrm{phot}}$, defined as,
\begin{equation}
z_{\mathrm{phot}}=\int z\,\mathrm{PDF}(z)\,\mathrm{d}z.
\label{avg_photoz}
\end{equation}
The corresponding uncertainty is assumed to be Gaussian and can be computed in a similar manner as the square root of the variance. When we apply this estimator to the PDFs from \texttt{ANNz2} we shall denote these mean redshifts estimates by AvgPDF-ANNz2. Note that the individual redshifts estimated directly from the \texttt{GPz} code already assume a Gaussian uncertainty, and so are already equivalent to Equation~\ref{avg_photoz}. As such we do not need to apply this estimate to the PDFs from \texttt{GPz}.

Secondly, we derive an estimate for the photometric redshift for each galaxy by summing the PDF to construct the cumulative distribution function (CDF), which we can randomly sample in a Monte-Carlo process.
This process consists in estimating the $z_{\mathrm{phot}}$ by using the image of a random number between $\left[0,1\right)$ for the inverse of the cumulative distribution function in each galaxy.
With this method we ensure that the redshift estimates are representative of the full underlying PDF information.
We expect that the distribution function of the single number redshift obtained through this method is equivalent to the stacked PDF of all galaxies in the data set. Moreover we await to reduce the systematic errors compared with any other photometric redshift estimator according to \citet{wittman}.

In summary, we have defined the following two pairs of photometric redshift estimators for this work: AvgPDF-ANNz2 and GPz (both assuming a Gaussian uncertainty); and CDF-ANNz2 and CDF-GPz (both estimated using the Monte-Carlo method).

\section{Results}
\label{results}

\begin{figure*}
 \centering
\subfigure{\includegraphics[width=\linewidth]{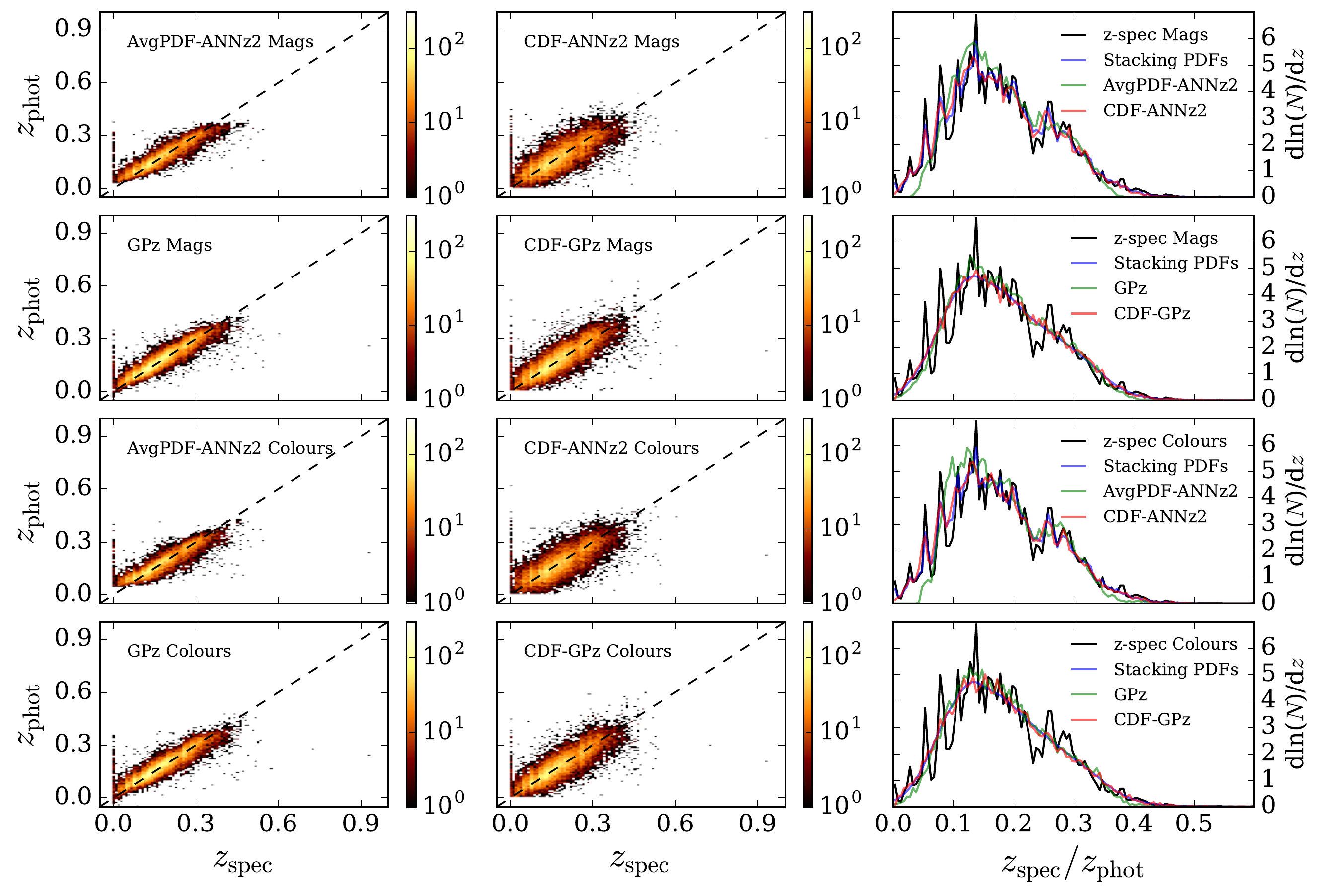}}
\subfigure{\includegraphics[width=\linewidth]{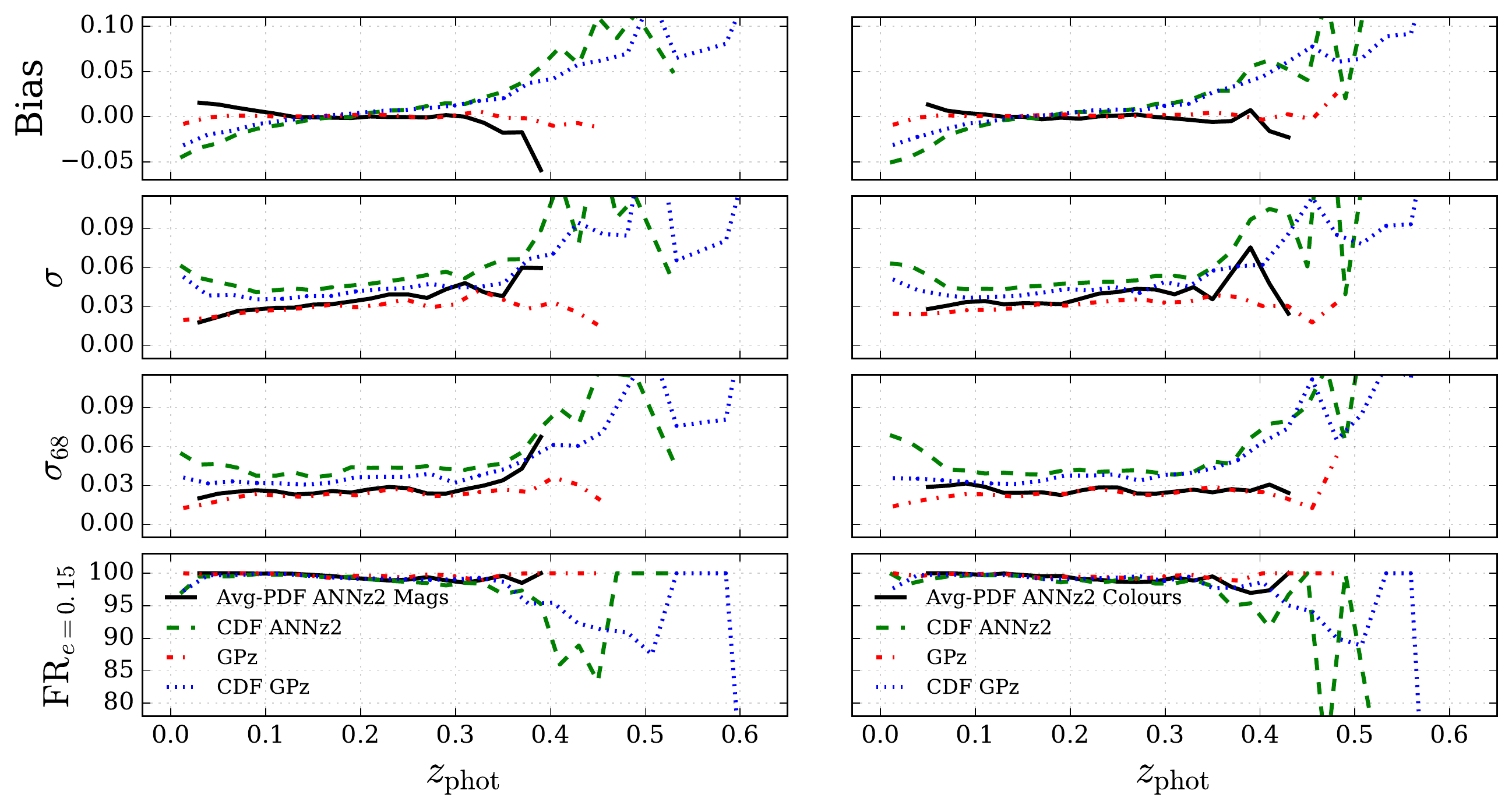}}
\caption{Statistical analysis of the photometric redshift computed for the real dataset by using magnitudes and colours.
\textbf{Top}: The two first columns are the scatter plots $z_{\mathrm{spec}}$ against $z_{\mathrm{phot}}$ for each photometric redshift estimator.
The last column are the $z_{\mathrm{spec}}$/$z_{\mathrm{phot}}$ distributions.
\textbf{Bottom}: Metrics as function of photometric redshift for each estimator (\textit{left}: magnitudes, \textit{right:} colours).} 
\label{zphotGamma}
\end{figure*}

\begin{figure*}
 \centering
\subfigure{\includegraphics[width=\linewidth]{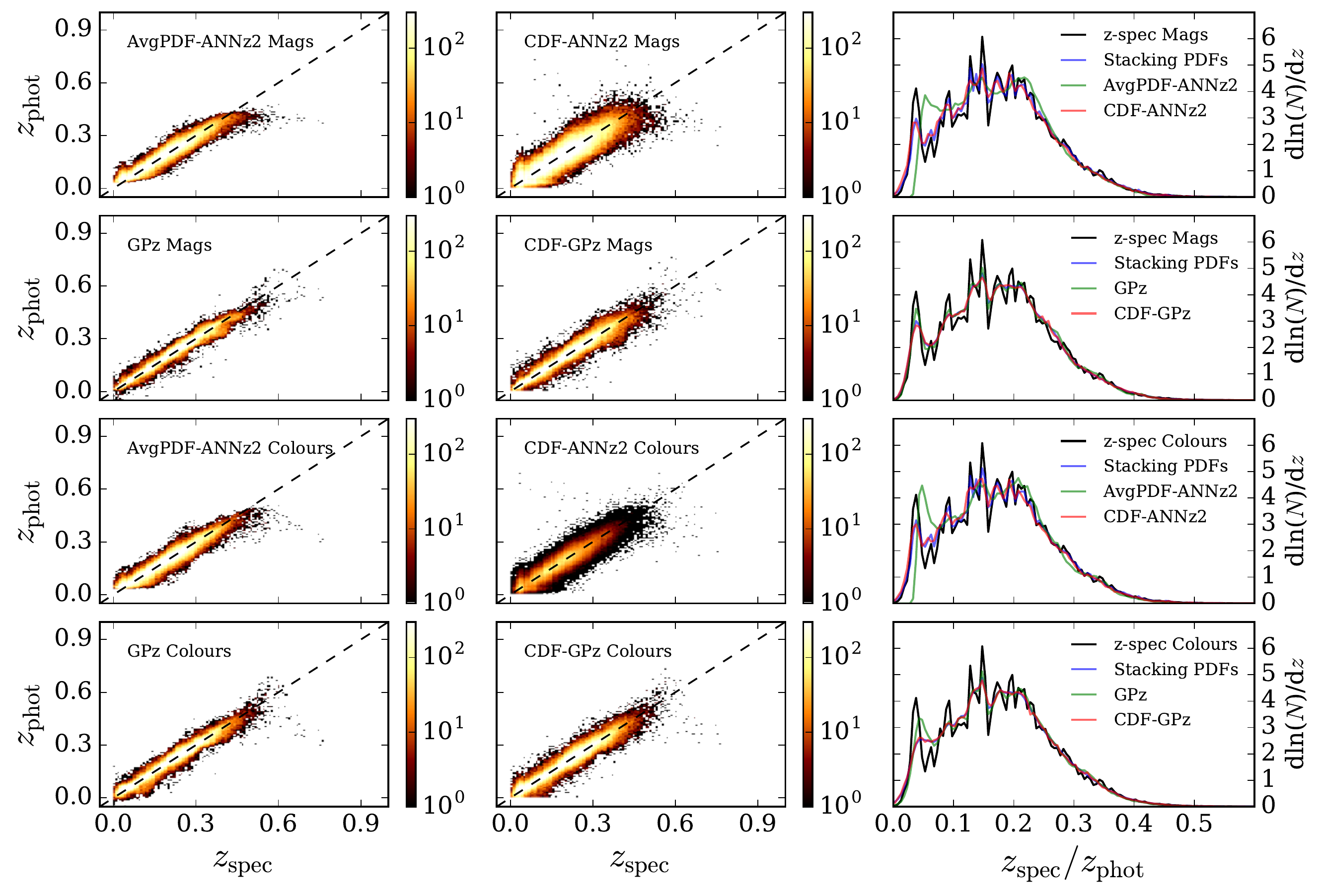}}
\subfigure{\includegraphics[width=\linewidth]{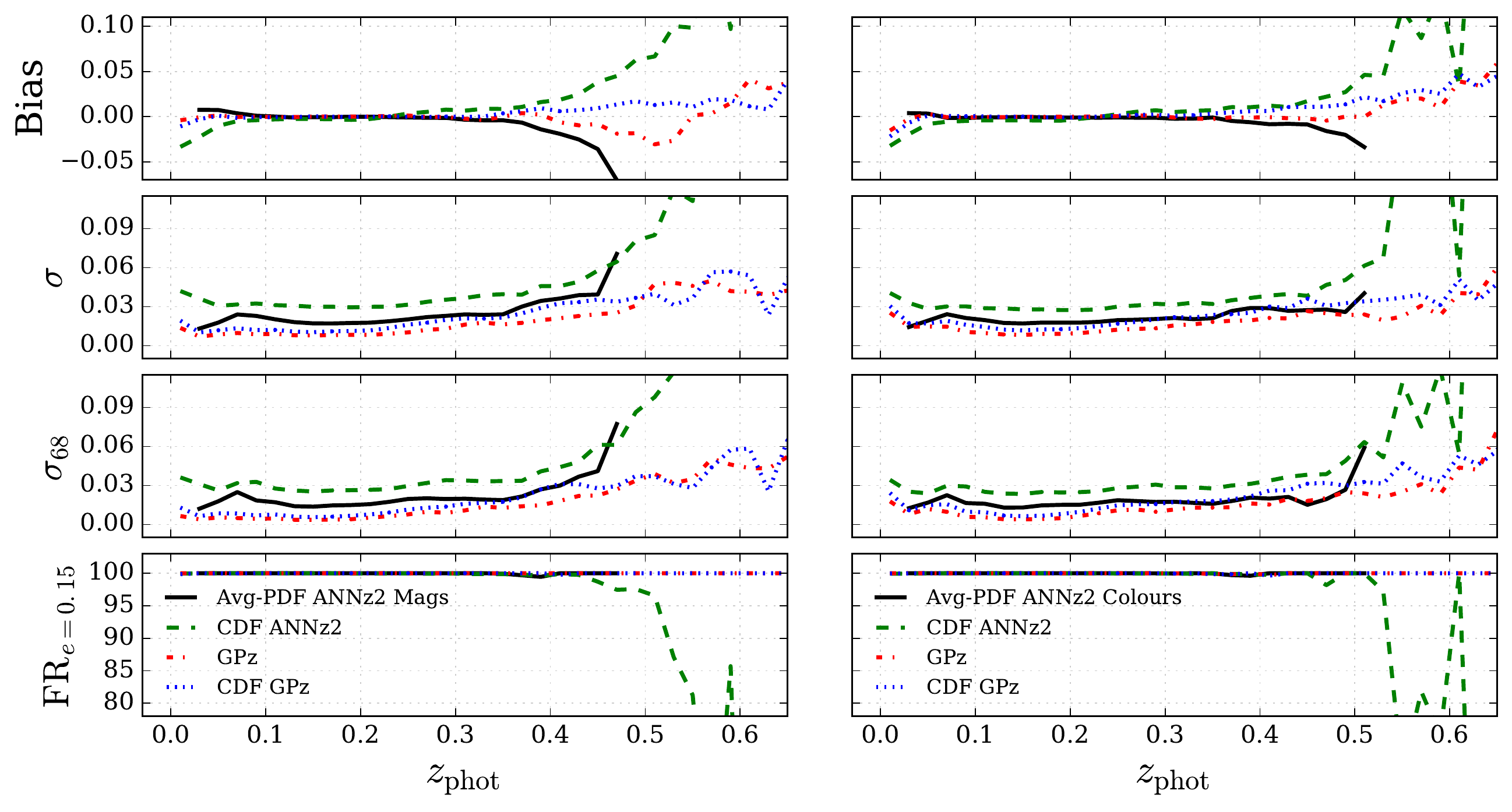}}
\caption{Statistical analysis of the photometric redshift computed for the mock catalogue by using magnitudes and colours.
\textbf{Top}: The two first columns are the scatter plots $z_{\mathrm{spec}}$ against $z_{\mathrm{phot}}$ for each photometric redshift estimator.
The last column are the $z_{\mathrm{spec}}$/$z_{\mathrm{phot}}$ distributions.
\textbf{Bottom}: Metrics as function of photometric redshift for each estimator (\textit{left}: magnitudes, \textit{right:} colours).} 
\label{zphotMock}
\end{figure*}

Initially in section~\ref{results_comparison} we compare the quality of the photometric redshifts obtained from our mock data and those from our real SDSS data. We then compare the quality of the photometric redshifts obtained when our real data is trained using a magnitude-selected training set and when the data is trained using a colour-selected training set.  Magnitude-selected training and colour-selected training are additionally applied to the mock catalogue.
Subsequently, in section~\ref{performance_degradation}, we select sequentially deeper $r$-band selected samples from the mock catalogue to analyse the degradation of the photometric redshifts recovered from each estimator when using a non-representative magnitude-selected training set.

\subsection{Comparison of real data and mock catalogue with a complete training set}
\label{results_comparison}

In this section, we apply both \texttt{ANNz2} and \texttt{GPz} codes to real data and mock catalogues with two different training choices, using the 5 SDSS magnitude bands in one case and 4 colours in the other.
Our aim is twofold: firstly, we want to confirm that our analysis with mock data is qualitatively consistent with the results we obtain in real data. 
Additionally, we wish to compare colour and magnitude types of training and assess their relative performance. 

When considering our real data, the GAMA MAIN data will be the testing set when training with a magnitude-selected training set and the GAMA DEEP data will be the testing set when training with a colour-selected training set. Details of the construction of the GAMA MAIN and GAMA DEEP samples, which we shall refer to collectively as the GAMA test data, are given in section~\ref{sdss_dr12}.
For the photometric analysis in the GAMA test data, we take \texttt{dered\_modelMag} (i.e. SDSS model magnitudes corrected for extinction) as the galaxy magnitudes and \texttt{modelMagErr} (Error in \texttt{modelMag}) as the magnitude errors. 
Since the magnitude limit in the spectroscopic GAMA dataset is $r_{\textrm{petro}}<19.4$ we apply a similar $r$-band cut to the mock catalogue, obtaining a mock training sample of 240 730 galaxies.
The \texttt{GPz} code provides us a function that allows us to split the spectroscopic GAMA sample and the mock catalogue in three subsamples: a training data set, a validation data set and a test data set, the last subsample is used to test the training in each case.
Table~\ref{size_samples} shows the number of galaxies in each subsample for both the real data and the mock data.

\begin{table}
\centering
\caption{Number of galaxies and the threshold $r$-band for every used subsample in the training of the real data and the mock data. We use the same $r$-band magnitude range for both datasets. 
See section~\ref{results_comparison} for details.}
\label{size_samples}
\begin{tabular}{c c c c c}
\hline \hline
Sample & Training & Validation & Testing & $r$-band range \\ \hline 
 GAMA & 20 864 & 20 865 & 21 497 & $r<19.4$ \\ 
 Mock & 20 220 & 20 222 & 200 288 & $r<19.4$ \\ \hline  
\end{tabular}
\end{table}

  The photometric analysis of the GAMA sample and mock catalogue sample is performed using the magnitudes ($u$, $g$, $r$, $i$, $z$) and colours ($u$-$g$, $g$-$r$, $r$-$i$, $i$-$z$).
For each colour $C(m_1,m_2)=m_2-m_1$ the error on the colour are obtained via standard error propagation:
\begin{equation}
\delta C(\delta m_1,\delta m_2)=\sqrt{\delta m_1^2+\delta m_2^2},
\end{equation}
where $\delta m_1$ and $\delta m_2$ are the errors on the magnitudes $m_1$ and $m_2$.

\begin{figure*}
 \centering 
 \includegraphics[width=\linewidth]{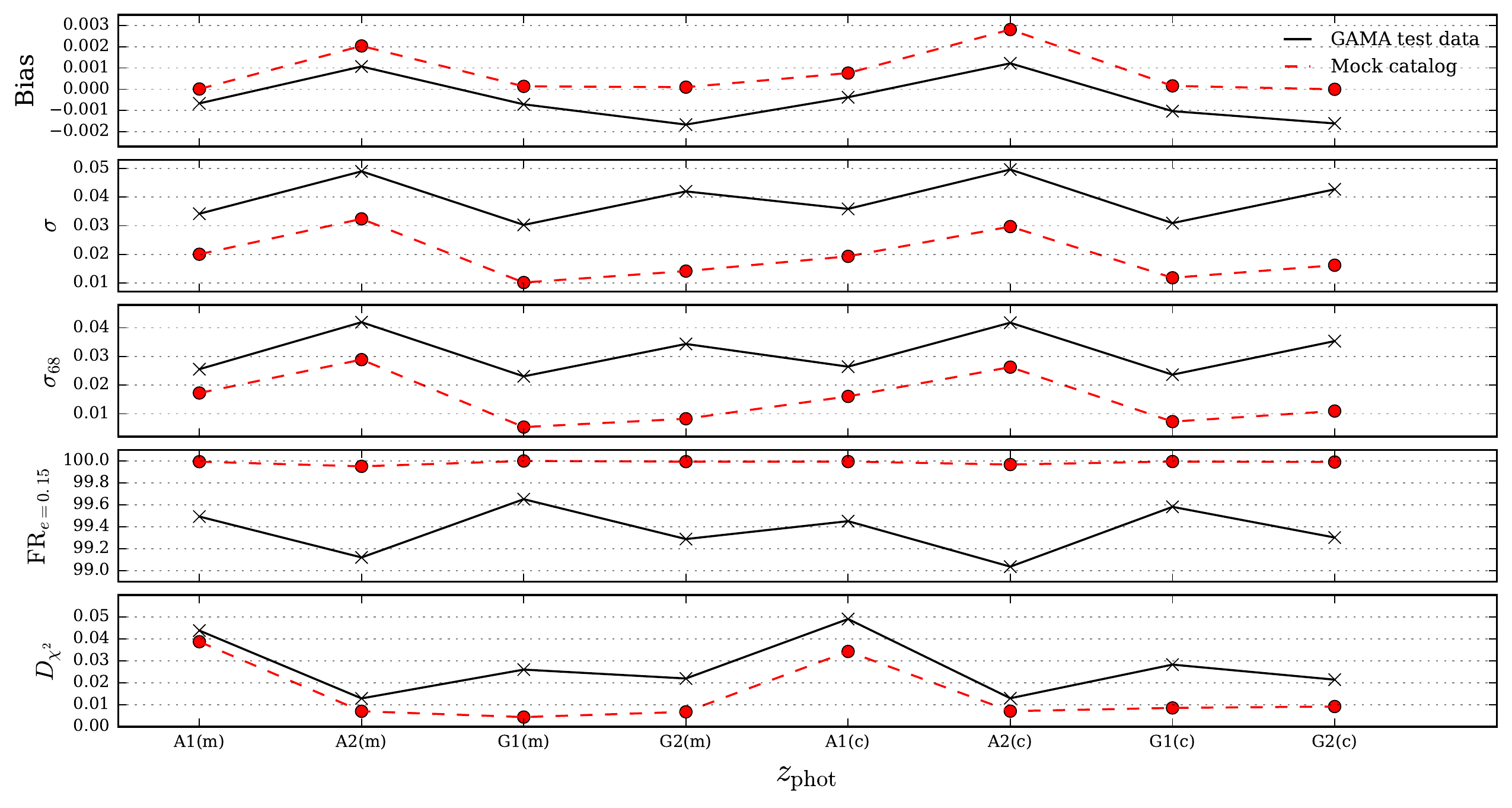}
 \caption{Comparison of global metrics for GAMA test data and the mock catalogue for each photometric redshift estimator by using magnitudes and colours. 
Here we use the following notation A1: AvgPDF-ANNz2, A2: CDF-ANNz2, G1: GPz, G2: CDF-GPz. The last letter indicates whether we compute the z-phot via the magnitudes (m) or colours (c). 
The bottom plot is the chi-square measure $(D_{\chi^2})$ which compares the distribution function for every estimator with the spectroscopic distribution function.} 
 \label{Comparison}
\end{figure*}

In the upper grids of panels in Fig.~\ref{zphotGamma} and Fig.~\ref{zphotMock} we compare for each photometric redshift estimator the recovered photometric redshifts to the spectroscopic redshifts of the galaxies. 
We show the results for both a magnitude-selected training and a colour-selected training of the mock catalogue and GAMA test data. 
We compare the corresponding redshift distribution functions. 
In the lower grid of panels we compare the metrics for each estimator. 
We see very little difference between training with a colour-selected training set and a magnitude-selected training set.
For both the real data and the mock catalogue, we note that the CDF cases show slightly more scatter compared to the AvgPDF-ANNz2 and GPz cases. 
However, when we examine the metrics we see that for both the mock catalogue and the GAMA data sets, over the redshift range $0.1 \lesssim z \lesssim 0.4$ each photometric redshift estimator yields a bias and a fraction of catastrophic outliers that is in excellent agreement with the other estimators, further there is a good agreement between the mock and the data. 

For the GAMA data sets, we see that compared to the two Gaussian estimators the CDF cases are typically able to estimate photometric redshifts out to higher redshifts beyond $z_{\rm phot}\sim0.4$, though these photometric redshifts have a larger scatter, a larger bias and a greater number of outliers (as indicated by a decreasing value for $\textrm{FR}_{e=0.15}$).
In the mock catalogue we observe the similar effect in the CDF-ANNz2 estimator compared with the AvgPDF-ANNz2 estimator. Moreover, the estimators based on Gaussian GPz are also able to recover photometric redshifts out beyond $z_{\mathrm{phot}}\sim0.4$.
The quantity $\textrm{FR}_{e=0.15}$ shows different trends in the data of the mocks, it deviates from 1 at $z\sim0.5$ in the mock and at $z_{\rm phot}\sim0.4$ on the data.
We also note that for the mocks, the $\textrm{FR}_{e=0.15}$ values for CDF-GPz and GPz estimators remain close to unity out to $z_{\rm phot}\sim0.65$.
Overall, however the agreement is excellent between the mock catalogue and real data.

We also note some sample variance features in both the mock and data (e.g. for real data, the estimators based on \texttt{ANNz2} code recover the peak in the data stack at $z\sim0.3$, unlike the GPz and CDF-GPz estimators).
These features disappear with some estimations.
The plots of the distribution functions show that those distributions based in the single value which are obtained through the Monte-Carlo method fits better with the stacking of galaxy PDFs than the AvgPDF-ANNz2 and GPz estimators.  
In fact, the previous assertion is more noticeable in the photometric redshifts estimated by the \texttt{ANNz2} code than in the photometric redshifts estimated by the \texttt{GPz} code. 
According to the chi-square measure presented in Fig.~\ref{Comparison}, for the GAMA case, the CDF-ANNz2 distribution fits better the spectroscopic redshift distribution than the distributions obtained through the other estimators.
In the case of the mock, the CDF-ANNz2, GPz and CDF-GPz estimators have similar chi-square measures and their distributions fit better the spectroscopic redshift distribution than the AvgPDF distribution. 

In Fig.~\ref{Comparison} we show for each photometric redshift estimator the global metric values (i.e. we compute the metric over all redshift) for both the GAMA test data and the mock catalogue. 
The global values are shown for both the magnitude-selected training and the colour-selected training.
In order to identify the cases and photometric redshift estimator used here, we employ the following notation AvgPDF-ANNz2 (A1), CDF-ANNz2 (A2), GPz (G1), CDF-GPz (G2). 
The final letter indicates whether we compute the photometric redshift via magnitude-selected training (m) or colour-selected training (c). 
Furthermore, the bottom panel shows the chi-square measure, given by Equation~\ref{chi_dist}), in each instance. 
We observe that the results obtained by using the magnitude-selected and colour-selected for the mock catalogue and the GAMA test data are similar. 
The scatter and the fraction $\textrm{FR}_{e=0.15}$ for the photometric redshifts in the mock catalogue are overall better than the equivalent metrics for the GAMA test data. 
It is clear that the mock catalogue is unable to properly model a $\sim0.5$ per cent catastrophic failure rate and have errors that are slightly too optimistic, though the mock catalogue has managed to simulate the overall qualities of the real data. 
We would expect larger photometric errors in the real data due to additional sources of error not included in the mock catalogues, such as the sky background on a given night or the effects of proximity to bright objects in the sky.
This similarity gives us confidence that these mock catalogues are suitable for examining the degradation in the next section. 
The results obtained from the mock catalogue show the same qualitative trends as the results for the GAMA test data, and we therefore claim that using the mock catalogue for the performance degradation analysis of the next section is suitable to show any degradation trends that would also be observed in real data.

\subsection{Performance degradation} 
\label{performance_degradation}

Having established the qualitative equivalence between the observed data and the mock catalogues, we will use the latter to evaluate the performance of the AvgPDF-ANNz2, CDF-ANNz2, GPz and CDF-GPz estimators when the training set is not representative in magnitude space.
The idea being that we can safely extrapolate a certain amount in the $r$-band magnitude given that we have a representative set in colour space. 
We construct several testing sets from the mock catalogue, by varying the $r$-band limiting magnitude in the range $\left[19.4,20.8\right]$ in steps of 0.2 magnitudes, i.e. with $\textrm{dm}_r=0.2$. 
Table~\ref{num_obj_rvary}  shows the number of objects for each sample used in this analysis.
For each testing set we use same the training and validation sets that were used to estimate the photometric redshifts for the previous mock catalogue analysis. 
These training and validation sets are selected from the mock catalogue with a magnitude cut of $r<19.4$. 
Since the training set is not representative in the magnitude space of the deeper testing sets, we work by using the colour-selected training (i.e. the colour space as input) to estimate the photometric redshifts. 
Our goal is to demonstrate that we can obtain reliable redshift distributions for fainter objects if we ensure representativeness in colour space. 
This can help to mitigate the impact of the non-representativeness problem in the training set of current large-scale structure surveys, where the available spectroscopic data sets are usually shallower than the overlapping photometric surveys. 
\begin{table}
\caption{Number of objects for each cut in the $r$-band magnitude.}\label{num_obj_rvary}
\centering
\begin{tabular}{c c }
\hline \hline
Cut of $r$-band & Number of objects \\ \hline 
 $r<19.4$& 200 288 \\ 
 $r<19.6$& 258 472 \\ 
 $r<19.8$& 330 181 \\ 
 $r<20.0$& 416 572 \\  
 $r<20.2$& 521 375 \\
 $r<20.4$& 647 349 \\
 $r<20.6$& 798 152 \\
 $r<20.8$& 978 533 \\ \hline 
\end{tabular}
\end{table}

\begin{figure*}
 \centering 
 \includegraphics[width=\linewidth]{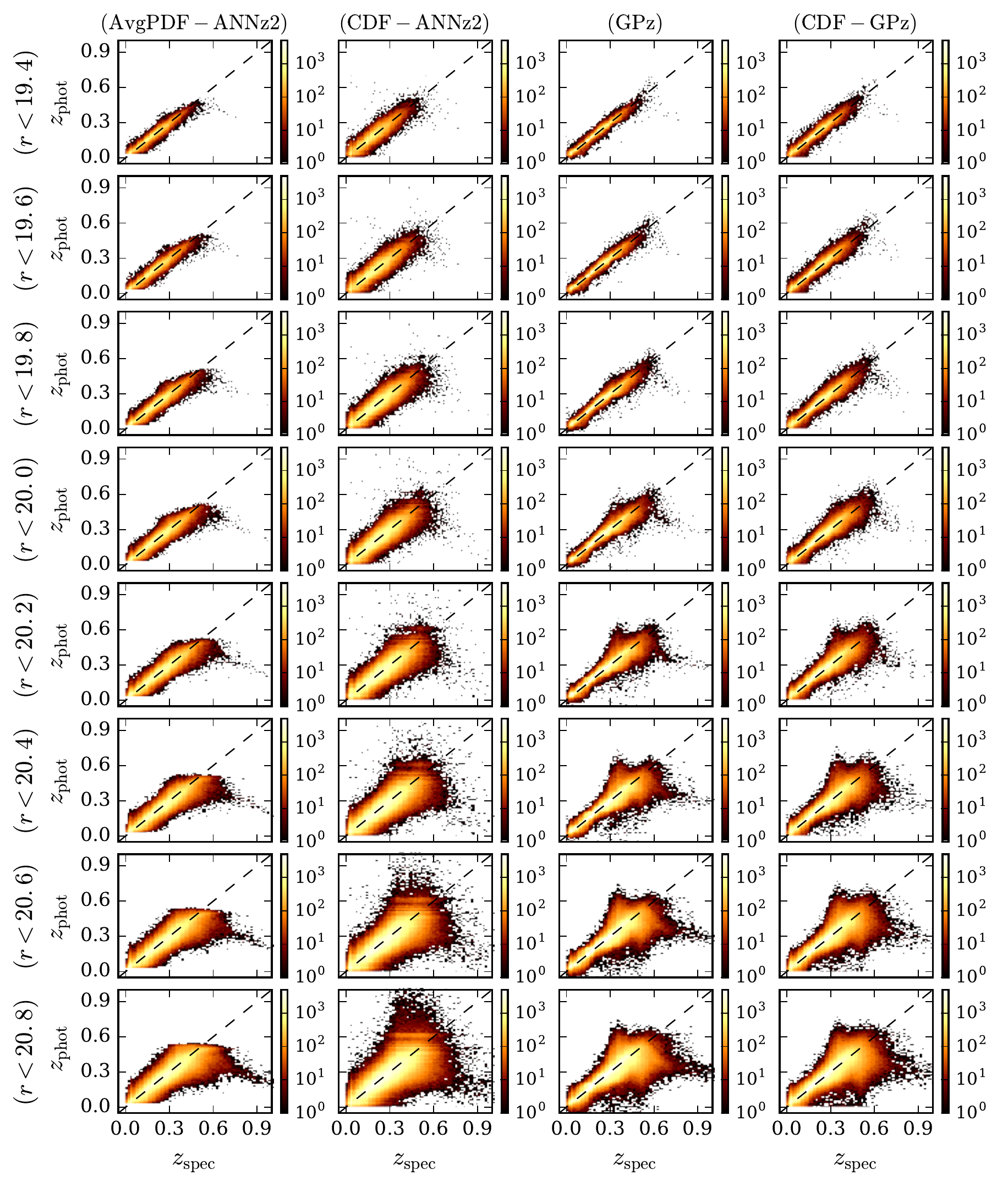}
 \caption{Scatter plots of z-spec against z-phot for $r$-band cuts in the range $\lbrack 19.4,20.8\rbrack$ by using the mock catalogue.
Here the colours are used as input for the photometric methods. 
The training set and validation  are obtained for $r<19.4$.
In the horizontal axis, we indicate the photometric redshift estimator used and in the vertical axis we indicate the $r$-band cut performed on mock catalogue.
Note that the scatter in the photometric redshift recovery increases with increasing magnitude depth for all methods. 
Moreover, for fainter flux limits the scatter increases with spectroscopic redshift.}
 \label{Scatter_all}
\end{figure*}

\begin{figure*}
 \centering 
 \includegraphics[width=\linewidth]{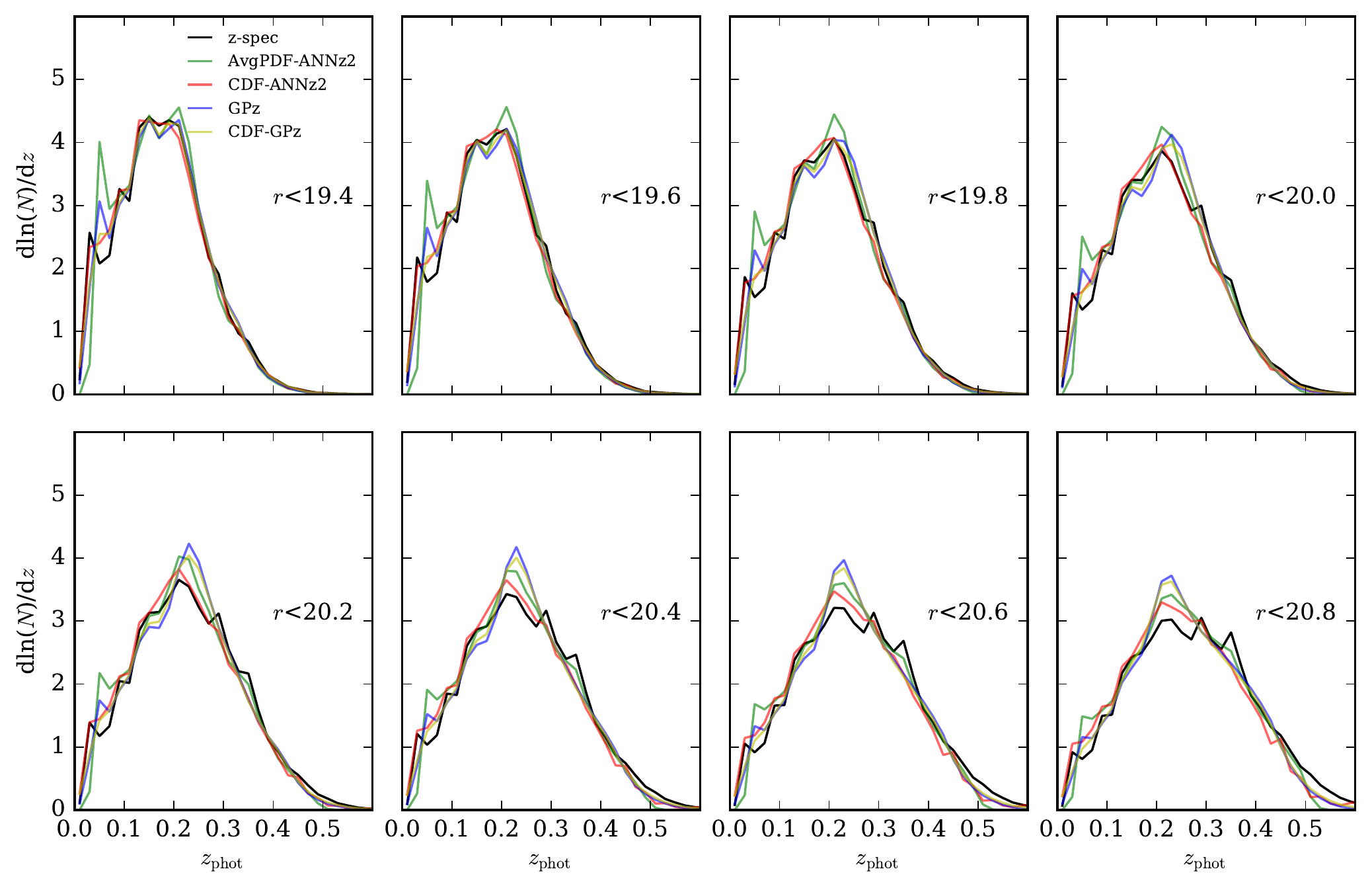}
 \caption{z-spec and redshift estimators distributions for $r$-band cuts in the range $\lbrack 19.4,20.8 \rbrack$ by using the mock catalogue.
Note that for fainter magnitude cuts (i.e., the cuts in the region $\left[20.2,20.8\right]$), the distributions of the estimated photometric redshifts have a peak in $z_{\mathrm{phot}} \approx 0.25$ which is not present in spectroscopic redshift distribution and a tail mismatch at higher $z$.
The effect is greater for the photometric redshift estimators obtained via the \texttt{GPz} algorithm.
In generally, the CDF-ANNz2 distribution fits better the z-spec distribution than the distributions obtained through the other estimators.}
\label{Dist_redshift}
\end{figure*}

We estimate the photometric redshifts by applying the same four estimators, as were used in the previous analysis, to the different $r$-band selected samples. 
In Fig.~\ref{Scatter_all} we plot, for each sample, the recovered photometric redshifts against the corresponding spectroscopic redshifts.
We note that the scatter in the photometric redshift recovery increases with increasing magnitude depth for all methods. 
Moreover, for fainter flux limits the scatter appears to increase with spectroscopic redshift. 
This is expected as fainter galaxies will have larger photometric errors and hence higher scatter in the photometric redshift space.
On the other hand, we can see that the AvgPDF-ANNz2 estimator is unable to recover photometric redshifts above $z_{\textrm{phot}}\gtrsim0.5$, an effect that worsens for fainter magnitude cuts. 
This is also expected due to the nature of the PDF fitting in \texttt{ANNz2} and the lack of training galaxies in the sample (i.e. the limited number of galaxies with $z_{\rm spec}>0.6$). 
Compared to the other three estimators the AvgPDF-ANNz2 estimator has a higher precision but low accuracy as we tend towards fainter magnitudes.
Note that the GPz and CDF-GPz estimators also struggle to recover many redshifts beyond $z_{\rm phot}\sim0.6$. 
Indeed for every estimator the one-to-one correspondence between spectroscopic and photometric redshift breaks down for redshifts above $z_{\rm phot}\sim0.6$. 
For $z\gtrsim0.6$ there is significant scatter and bias in the recovered redshifts, particularly in the samples with fainter magnitude selection.
In Fig.~\ref{Dist_redshift} we show the redshift distribution functions, as a function of limiting magnitude, for each of the photometric redshift estimators that we consider.   
We note that for every magnitude limit the CDF-ANNz2 estimator provides a better fit to the spectroscopic redshift distribution than the other photometric redshift estimators. 
The distributions from the two GPz estimators show good fit with the spectroscopic redshift distribution for all $r$-band cuts brighter than $r<20.0$. 
For fainter magnitude cuts, the distributions of the estimated photometric redshifts have a peak in $z_{\mathrm{phot}} \approx 0.25$ which is not present in the spectroscopic redshift distribution.
This peak comes hand in hand with a mismatch at higher redshift. 
The effect is more prominent for the two \texttt{GPz}-based photometric redshift estimators.
This peak excess is caused by galaxies that are identified with deeper magnitude selection and have a large spectroscopic redshift, but are estimated to have a smaller photometric redshift. 
These galaxies can be seen in the lower panels of Fig.~\ref{Scatter_all} as a long tail extending to high spectroscopic redshift.
We conclude that the photometric redshift distributions are very similar for magnitude limits brighter than $r<20.0$. 
Then, by using colour-selected training set, we manage to recover the true redshift distribution and estimate reliable photometric redshifts for $\sim 42\%$ of galaxies with respect to the deeper $r$-band cut whereas that testing set in which the magnitude-selected training set is representative only constitutes $\sim 20\%$.
In Appendix~\ref{metrics_depends_on_time} we present the metrics as function of each photometric redshift estimator for all $r$-band cuts.
These values allow the reader to have a better understanding of the discussion presented in this section.

\begin{figure*}
 \centering 
 \includegraphics[width=\linewidth]{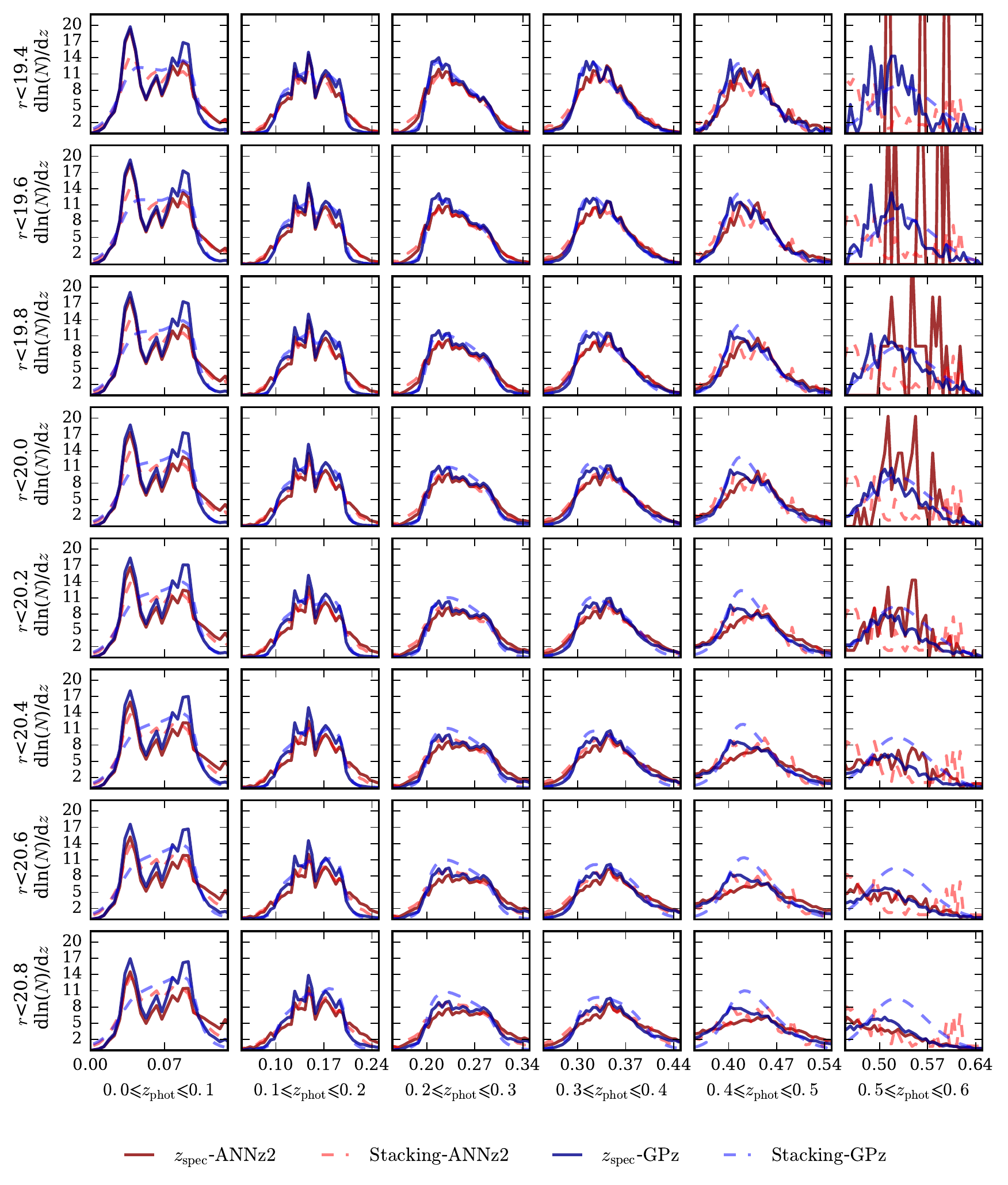}
 \caption{Comparison between the spectroscopic distribution and the stacking of photometric redshift PDFs estimated through \texttt{ANNz2} and \texttt{GPz} algorithms for slices of photometric redshift in all $r$-band magnitude cuts.
We estimate different spectroscopic redshift distribution for each used algorithm, since that the population of galaxies in the slices is different for each photometric redshift estimator.
Note that the stacking of photometric redshift computed with \texttt{ANNz2} algorithm fits better the spectroscopic distribution than the \texttt{GPz} case. 
Here we use black solid line for z-spec (ANNz2), red dashed line for stacking (ANNz2), blue solid line for z-spec (GPz) and green dashed line for stacking (GPz).}
 \label{Dist_bin_redshift}
\end{figure*}

In cosmological measurements with photometric large-scale structure surveys, much of the information is obtained by splitting the galaxy sample in several photometric redshift bins in order to measure auto- and cross-correlations between the sub-samples in the different bins.  
We are therefore interested in assessing the accuracy of the recovery of the redshift distribution in differential redshift bins. 
In Fig.~\ref{Dist_bin_redshift} we compare the stacking of the photometric redshift PDFs estimated through \texttt{ANNz2} and \texttt{GPz} codes with the spectroscopic distribution for slices of photometric redshift in all $r$-band magnitude cuts. 
We consider six photometric redshift bins of width $\mathrm{d}z_{\mathrm{phot}}=0.1$ between $0.0\leq z_{\mathrm{phot}}\leq0.6$.
The selection of galaxies in each redshift slice is performed with the AvgPDF-ANNz2 estimator for the \texttt{ANNz2} case and with the GPz estimator for the \texttt{GPz} case. 
Since the specific choice of galaxies in the slices is different for each photometric redshift estimator, we compute the spectroscopic redshift distribution associated to each algorithm. 
We observe that the stacking of photometric redshift PDFs computed with the \texttt{ANNz2} algorithm fits better the spectroscopic distribution than the \texttt{GPz} case. 
In the redshift bins within the range $0.1\leqslant z_{\rm phot} \leqslant 0.4$, there is good agreement between the stacking for both algorithms and the spectroscopic redshift distribution. 
However, for deeper magnitude selection this agreement worsens for both cases \texttt{ANNz2} and \texttt{GPz}. 
The stacking (GPz) presents the greatest differences with the spectroscopic redshift distribution in the redshift slices $0.4\leqslant z \leqslant 0.6$. 

\begin{figure*}
 \centering 
 \includegraphics[width=\linewidth]{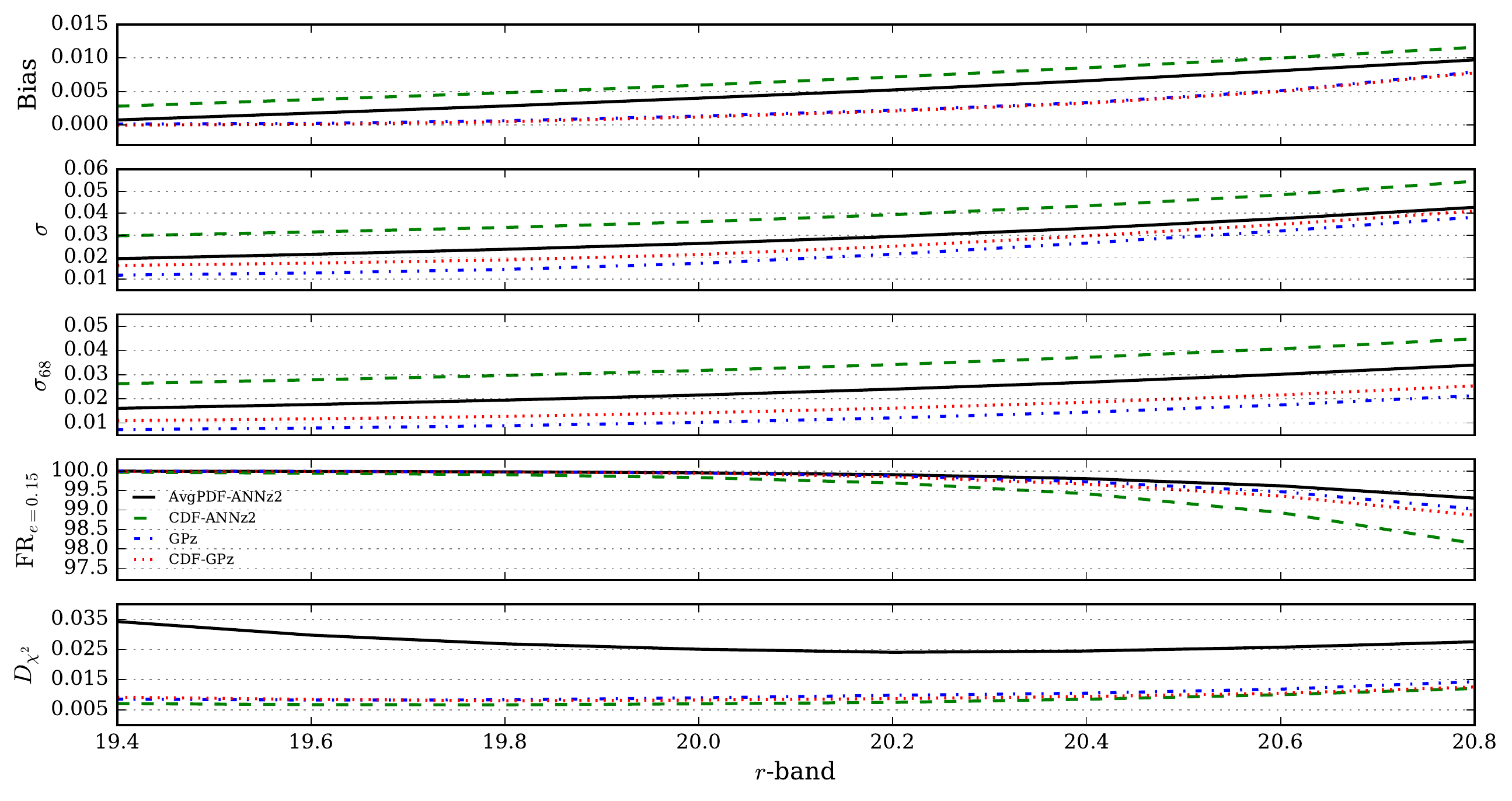}
 \caption{Global metrics of each photometric redshift estimator as function of $r$-band cut.
The last figure is the chi-square measure $(D_{\chi^2})$ for the $z_{\mathrm{spec}}$ distribution and $z_{\mathrm{phot}}$ one.
We observe that the CDF-ANNz2 estimator has the best chi-square measure in all $r$-band cuts.
Moreover, we note that the GPz estimators present the lowest global scatter and bias.}
 \label{Global_metrics}
\end{figure*}

In Fig.~\ref{Global_metrics} we show the global metric values and the chi-square measure for all redshift range of the samples as function of $r$-band cut. 
The metric values worsen towards deeper magnitude limits, as we might expect. 
However, for each of the photometric redshift estimators the fraction $\textrm{FR}_{e=0.15}$ remains above 99.5 per cent until $r\approx 20.2$. 
The AvgPDF-ANNz2 and CDF-ANNz2 estimators have the highest scatter and bias, though the CDF-ANNz2 estimator has the best chi-square measure for all $r$-band cuts. 
The GPz and CDF-GPz estimators present the lowest global scatter and bias, as well as high values for the global fraction $\textrm{FR}_{e=0.15}$. 
These estimators also have a low chi-square measure. 

In this subsection, the focus has been on comparing the different redshift runs. 
But for the science applications of these results, the important point is that for the best of the estimators (CDF-ANNz2), we can push the magnitude limit to a deeper range, and the degradation of redshift performance is only gradual. 
The performed test in slices of redshift shows us that the \texttt{ANNz2} code achieve good results in high redshifts for fainter magnitude cuts unlike to \texttt{GPz} code.  
Note that Monte-Carlo sampling of the PDF allows us to improve the accuracy of the photometric redshift values if we know the full photometric redshift PDF for every galaxy in the survey, as is the case when working with the \texttt{ANNz2}. 

\section{Implication for detection of galaxy clusters}
\label{implication_galaxy_clusters}

The reduced cost of measuring photometric redshifts, compared to spectroscopic redshifts, means that we are able to obtain photometric redshifts for many more objects more rapidly. 
As such, we can have large photometric galaxy surveys, which is statistically beneficial for many cosmological analyses, albeit with a reduced redshift precision compared to spectroscopic surveys. 
One such analysis is galaxy cluster counts.
Galaxy clusters are statistically very rare objects, at the extreme high mass end of the halo mass function, and so to maximise counts we need to probe large volumes.
On the other hand, for the detection of galaxy clusters we need to ascertain with as great an accuracy as possible which galaxies are members of the cluster and which are not. 
For this we need as accurate and precise redshift measurements as possible. 
Furthermore, to measure the halo mass function we need to estimate the halo mass of clusters.
One way is to estimate the mass dynamically, for which we need to accurately measure the positions of the cluster members to high precision \citep[see][]{borgani_a,borgani_b, voit,allen,kravtsov}. 
Therefore it is very important to estimate the photometric redshifts with accuracy and precision in order to minimise the impact on the systematic errors in the estimated number cluster count and subsequent cosmological analysis.
The main aim here is to examine the impact that using a non-representative training data set for photometric redshift estimation has on galaxy cluster detection with methods that are sensitive to the density of galaxies in a field, such as Voronoi Tessellation (VT) or kernel density estimation \citep[see][]{gal,lopes,soares}.
We also want to examine the impact of using each photometric redshift estimators used in this work.
We must remark that this analysis is not appropriate for cluster detection methods that are based upon dynamical measurements (e.g. Friends of Friends) or colour selection, which do not make use of density field information.

In redshift regions with higher density of galaxies we expect to find more galaxy clusters. 
Therefore in order to estimate the number of galaxy clusters that we can detect with a given redshift survey, we first compute the the number density of galaxies as a function of redshift, $n(z)$. 
This is equal to the number of galaxies, $N$, per comoving volume, $V$, and given by,
\begin{equation}
n(z)=\frac{\mathrm{d}N}{\textrm{d}V}=\frac{\mathrm{d}N}{\mathrm{d}z}\frac{\mathrm{d}z}{\mathrm{d}V} = \frac{\mathrm{d}N}{\mathrm{d}z}f_c(z),
\end{equation}
where
\begin{equation}
\label{num_per_vol_2}
f_c(z)\equiv\frac{H(z)}{D_c^2(z)\Delta\Omega}.
\end{equation}
Here $\mathrm{d}N/\mathrm{d}z$ corresponds to the galaxy redshift distribution, $\Delta\Omega$ is the angular area that the galaxy catalogue covers, $H(z)$ is the Hubble parameter and $D_c$ is the comoving distance.

\begin{figure*}
 \centering 
 \includegraphics[width=\linewidth]{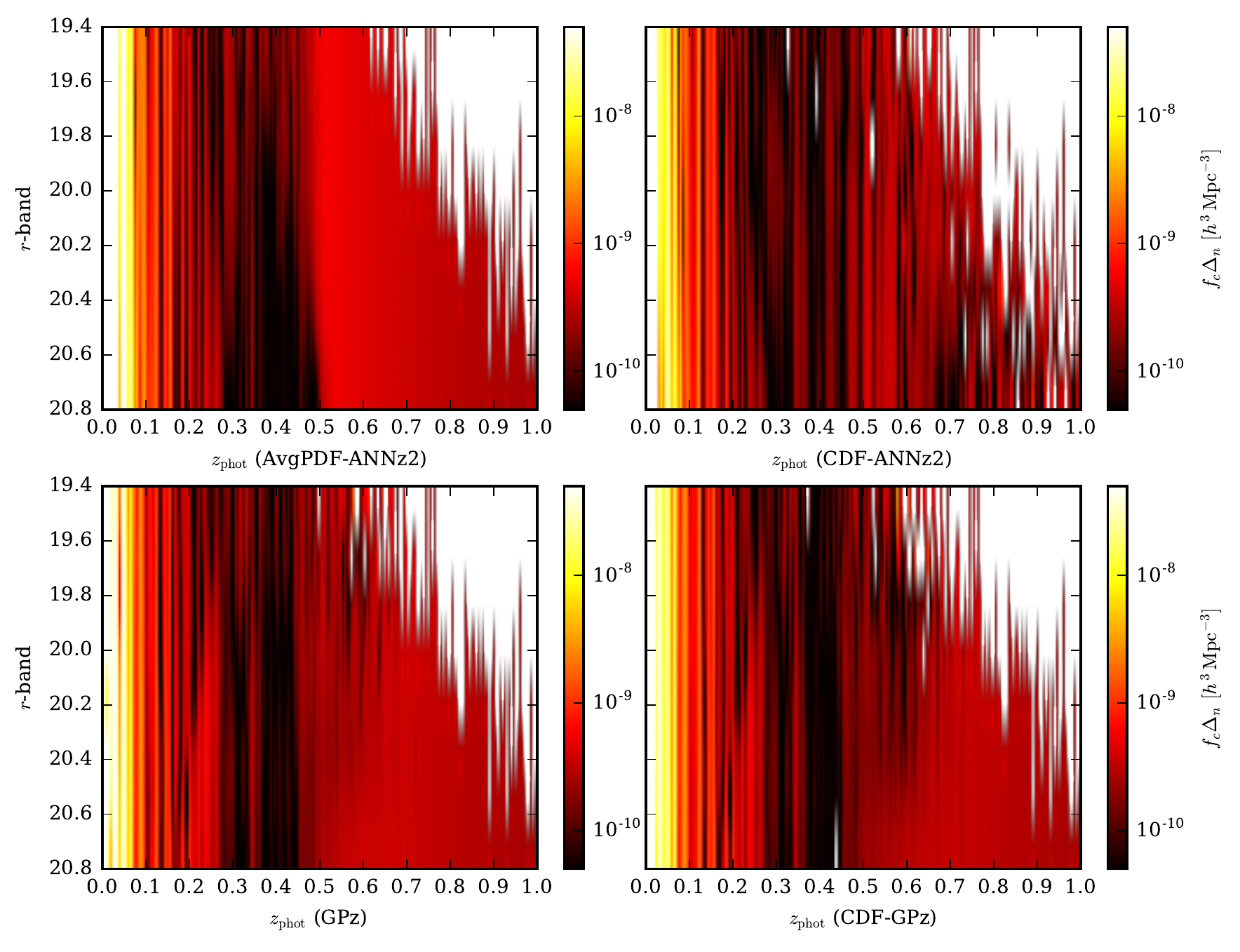}
 \caption{Comparison of number of galaxies per comoving volume element computed by using each photometric redshift estimator with the spectroscopic redshift case.
We compute the relative error $\Delta_n$ times the function $f_c(z)$ (this function contains the cosmological information, see Equation~(\ref{num_per_vol_2})) between density of galaxies for z-spec and the density of galaxies for each redshift estimator. This process is performed for all $r$-band cuts.
Note that the CDF-ANNz2 estimator allows us to detect galaxy clusters agreement with the z-spec data for deeper cuts in the $r$-band magnitude and highest redshifts, hence we expect that the galaxy cluster catalogue obtained by employing this photometric redshift estimator is purer until high redshifts than in other cases. 
}
 \label{GlobalCluster}
\end{figure*}

We compare the density of galaxies estimated using the photometric redshift distribution for each photometric redshift estimator with the density of galaxies estimated using the spectroscopic redshift. 
We make this comparison for each of our $r$-band magnitude cuts.
To quantify this comparison, we use the function $f_c(z)$ times the relative error between the two number densities, thus we have
\begin{equation}
\label{error_clusters}
f_c\Delta_n\equiv\frac{H}{D_a^2\Delta\Omega}\left|1-\frac{n}{\bar{n}}\right|,
\end{equation}
where $n$ is the number density of galaxies from the photometric redshift estimators and $\bar{n}$ is the number density of galaxies from the spectroscopic redshift.
This quantity is relevant as we would like to have a cluster detection method based on density estimation which is not affected by detection in the $n(z)$ function inferring incorrectly a different density of galaxies at that redshift.
Note that this calculation is not applicable to colour based methods to finding galaxy clusters, because here we are using the spatial information. 

In Fig.~\ref{GlobalCluster} we show the quantity $f_c\Delta_n$, described in Equation~(\ref{error_clusters}), where in each of the different panels we have used the photometric redshifts from the AvgPDF-ANNz2, CDF-ANNz2, GPz or CDF-GPz estimator.
In each panel darker colours correspond to smaller values for $f_c\Delta_n$, which indicates regions in the magnitude versus redshift space where the number densities derived from photometric redshifts are equal to the number densities from spectroscopic redshifts. 
Therefore in such regions we could robustly detect a galaxy cluster using both spectroscopic and photometric redshifts.
Note that for the CDF-ANNz2 estimator we see more darker regions at higher redshifts suggesting that with this estimator we can more robustly detect galaxy clusters at higher redshift with deeper $r$-band selected samples. 
Hence we would expect that a galaxy cluster catalogue obtained with this photometric redshift estimator would be purer, out to higher redshift, compared to catalogues build with the other estimators.
In other words, this result suggests that of all of the photometric redshift estimators considered, the CDF-ANNz2 estimator would provide the most accurate detection of galaxy clusters. 
The AvgPDF-ANNz2 estimator has the best results in the region $z \sim \,\left[0.25,0.50\right]$ and deeper $r$-band magnitude cuts. 
The GPz estimators have good results for the brighter magnitude cuts. 
However, for magnitude cuts fainter than $r<20.0$, we observe that in the redshift range $z \sim\,\left[0.2,0.3\right]$, the \texttt{GPz}-based estimators have larger values for $f_c\Delta_n$ than the \texttt{ANNz2}-based estimators (i.e. the \texttt{GPz}-based estimators have fewer darker regions than the \texttt{ANNz2}-based estimators). 
This is understandable as this within this redshift range where, in the lower panels of Fig.~\ref{Dist_redshift}, we saw a spurious peak in the photometric redshifts from the \texttt{GPz}-based estimators.
We conclude that the results presented here can be used to guide parameter optimisation of cluster finding algorithms.

\section{Conclusions and remarks}
\label{conclusions}
Photometric redshifts allow us to probe much larger volumes of the Universe than it is possible with spectroscopic redshifts, but they have large measurement uncertainties. 
Machine learning methods are often used to estimate photometric redshifts, but these estimators must be trained using existing spectroscopically detected datasets, which probe a limited volume. 
There is much uncertainty regarding the reliability of measured photometric redshifts when the spectroscopic training set is not representative of the  photometric dataset. 
In this work we have investigated the degradation in the accuracy and precision of the recovered of photometric redshifts when two machine learning methods, applied to deep photometric datasets, are trained using much shallower and brighter spectroscopic samples. 
We have used the \texttt{ANNz2} and \texttt{GPz} machine learning codes for estimating the photometric redshifts with four colours instead of all five magnitudes as input, ensuring representativeness only in this subspace, and evaluated the consequences on mock catalogues.
For this analysis, we also utilise Monte-Carlo random sampling for estimating a photometric redshift based on the full information in the cumulative distribution function (CDF) of the redshift probability distribution function (PDF). 
Altogether we use four photometric redshift estimators in this work; AvgPDF-ANNz2, CDF-ANNz2, GPz and CDF-GPz, which we define and introduce in section~\ref{photometric_estimators}.
In order to measure the quality of the estimated photometric redshifts we use the following typical metrics: Bias, $\sigma$, $\sigma_{68}$ and level of outliers $\textrm{FR}_e$ for a outlier threshold equal to $e=0.15$, see section~\ref{Metrics_work}. 

We start by showing that, for a representative training data set in the magnitude space, the photometric redshifts obtained using the \texttt{ANNz2} and \texttt{GPz}  algorithms display similar quality, either using magnitude-selected or colour-selected training sets as input.
We estimate the photometric redshift for the samples GAMA DEEP and GAMA MAIN (subsamples from the SDSS DR12 data with GAMA selections, see section~\ref{sdss_dr12}), which are trained by the spectroscopic GAMA survey.
In general, we find that the results in the metrics obtained for the mock catalogue display similar trends to the results metrics obtained for the GAMA test data. 
We observe that the photometric redshift distribution obtained with the CDF-ANNz2 estimator is the most consistent with the spectroscopic redshift distribution for the GAMA test data. 
We note that the distribution of the photometric redshifts obtained with those estimators that sample the CDF are a better fit to the photometric redshift PDF stacking of all galaxies in the data set. 
Nonetheless, these estimators yield a greater scatter than the other estimators. 

We proceed to analyse samples of the mock catalogue selected using progressively deeper cuts in the $r$-band magnitude in order to study the degradation of the photometric redshifts  obtained from the AvgPDF-ANNz2, CDF-ANNz2, GPz and CDF-GPz estimators when the training data set is non-representative of a deeper photometric testing set. 
In each instance we use the same training data set selected with $r<19.4$.
The AvgPDF-ANNz2 estimator does not recover high redshift and this fact worsens for deeper cuts. 
We consider that this result is due to the low density of spectroscopic galaxies at high redshift in the training set. 
Comparatively, the CDF-ANNz2 estimator shows better performance at higher redshifts, albeit with larger scatter. 
We observe that the CDF-ANNz2 estimator has the best chi-square measure for all $r$-band selections. 
The GPz and CDF-GPz estimators, appear to provide more reliable results at low redshifts. 
Nevertheless for deeper cuts, we observe that these estimators tend to under-estimate the redshifts of high-redshift spectroscopic galaxies leading to an excess of photometric redshifts at the peak of the redshift distribution and a mismatch in the tail of the distribution. 
For the scatter plots between spectroscopic redshifts and photometric redshifts as well as the $n(z)$ plots up to $r<20.0$ we observe very good results in all photometric redshift estimators.
The colour-selected training set allows us to estimate reliable photometric redshifts 
for a testing data set two time as large as testing data set in which the magnitude-selected training set is representative, see Figure~\ref{Scatter_all} and Figure~\ref{Dist_redshift}. 

The large surveys of photometric redshift offer us an excellent tool to perform cosmological analysis, in particularly, abundance of galaxy clusters.
In the last section, we are informing galaxy cluster searches by highlighting regions where robust cluster detection is more likely (i.e. denser regions).
In order to quantify the impact of the photometric redshifts in the detection of galaxy clusters, we compute the number of galaxies per comoving volume for each single value redshift estimator. 
This quantity provides us information about clusters detected via methods based on density of galaxies, such as VT or kernel density estimation. 
To compute the number density of galaxies, we use the photometric redshift distribution obtained for each estimator.
Note that the depth of the $r$-band magnitude cut is directly related with the density of galaxies and hence the number of galaxy clusters detected.
However, we must recall that the estimated photometric redshifts become poorer quality for deeper cuts.
The density of galaxies given by CDF-ANNz2 estimator has the best agreement with the number density of galaxies given by spectroscopic redshift data in deeper cuts and high redshifts.
For lower redshifts and $r$-band magnitude cuts, the number density based on the other estimators have better agreement with the spectroscopic number density, nonetheless the number density based on the CDF-ANNz2 estimator also has good agreement, see Figure~\ref{GlobalCluster}. 
We conclude that the results here can improve detectability of clusters with density based detection methods.

\section*{Acknowledgements}

JDR acknowledges Coordena\c c\~ao de aperfei\c coamento de pessoal de nivel superior (CAPES) for financial support. MCBA acknowledges to Conselho Nacional de Desenvolvimento Cient\'ifico e Tecnol\'ogico (CNPq) for partial support. 
This work was supported by the Science and Technology Facilities Council [ST/J501013/1, ST/L00075X/1]. This work used the DiRAC Data Centric system at Durham University, operated by the Institute for Computational Cosmology on behalf of the STFC DiRAC HPC Facility www.dirac.ac.uk). This equipment was funded by BIS National E-infrastructure capital grant ST/K00042X/1, STFC capital grant ST/H008519/1, and STFC DiRAC Operations grant ST/K003267/1 and Durham University. DiRAC is part of the National E-Infrastructure.
FBA acknowledges the support of the Royal Society via a RSURF.








\appendix

\section{GAMA MAIN and GAMA DEEP SQL query}
\label{SQL_query}

In order to build the GAMA MAIN and GAMA DEEP samples from the SDSS DR12 database, we used the following SQL query:

\begin{verbatim}
SELECT
  [a selection of parameters from SDSS tables]
FROM
  PhotoPrimary AS p
  JOIN Field AS f ON f.fieldID = p.fieldID
  JOIN Run AS r ON f.run = r.run
WHERE
--/ QUALITY FLAGS
  ((p.calibStatus_g & 1) != 0) 
  AND ((p.calibStatus_r & 1) != 0) 
  AND ((p.calibStatus_i & 1) != 0)
  AND p.cModelMag_i < 21.0
  AND ((p.flags_g & 0x80000000802) = 0)
  AND (((p.flags_g & 0x8) = 0) 
  OR ((p.flags_g & 0x40) = 0)) 
  AND ((p.flags_r & 0x80000000802) = 0)
  AND (((p.flags_r & 0x8) = 0) 
  OR ((p.flags_r & 0x40) = 0)) 
  AND ((p.flags_i & 0x80000000802) = 0)
  AND (((p.flags_i & 0x8) = 0) 
  OR ((p.flags_i & 0x40) = 0))
--/ STAR-GALAXY SEPARATION
  AND p.type = 3
  AND ((p.psfMag_r - p.modelMag_r) > 0.25)
--/ GAMA CUTS
  AND ((p.modelMag_u - p.modelMag_g) > -2) 
  AND ((p.modelMag_u - p.modelMag_g) < 7)
  AND ((p.modelMag_g - p.modelMag_r) > -2) 
  AND ((p.modelMag_g - p.modelMag_r) < 5)
  AND ((p.modelMag_r - p.modelMag_i) > -2) 
  AND ((p.modelMag_r - p.modelMag_i) < 5)
  AND ((p.modelMag_i - p.modelMag_z) > -2) 
  AND ((p.modelMag_i - p.modelMag_z) < 5)
  AND p.petroMag_r > 12
  AND p.petroMag_r < 19.4 --\ only for GAMA MAIN
\end{verbatim}

\section{Metrics depending on redshift}
\label{metrics_depends_on_time}

The scatter plot in Fig.~\ref{Scatter_all} and the distribution function of redshift in Fig.~\ref{Dist_redshift} allow us to describe the accuracy and the precision for each $r$-band cut employed in this work. 
However, to quantify these propierties we can computed the metrics described in the section~\ref{Metrics_work}. 
In Fig.~\ref{Global_metrics} we show the global metrics (i.e. the metric value for all redshift range) for all tests.  
In this appendix we present the metrics as function of photometric redshift estimator for each $r$-band cut. 

\begin{figure*}
 \centering 
 \includegraphics[width=\linewidth]{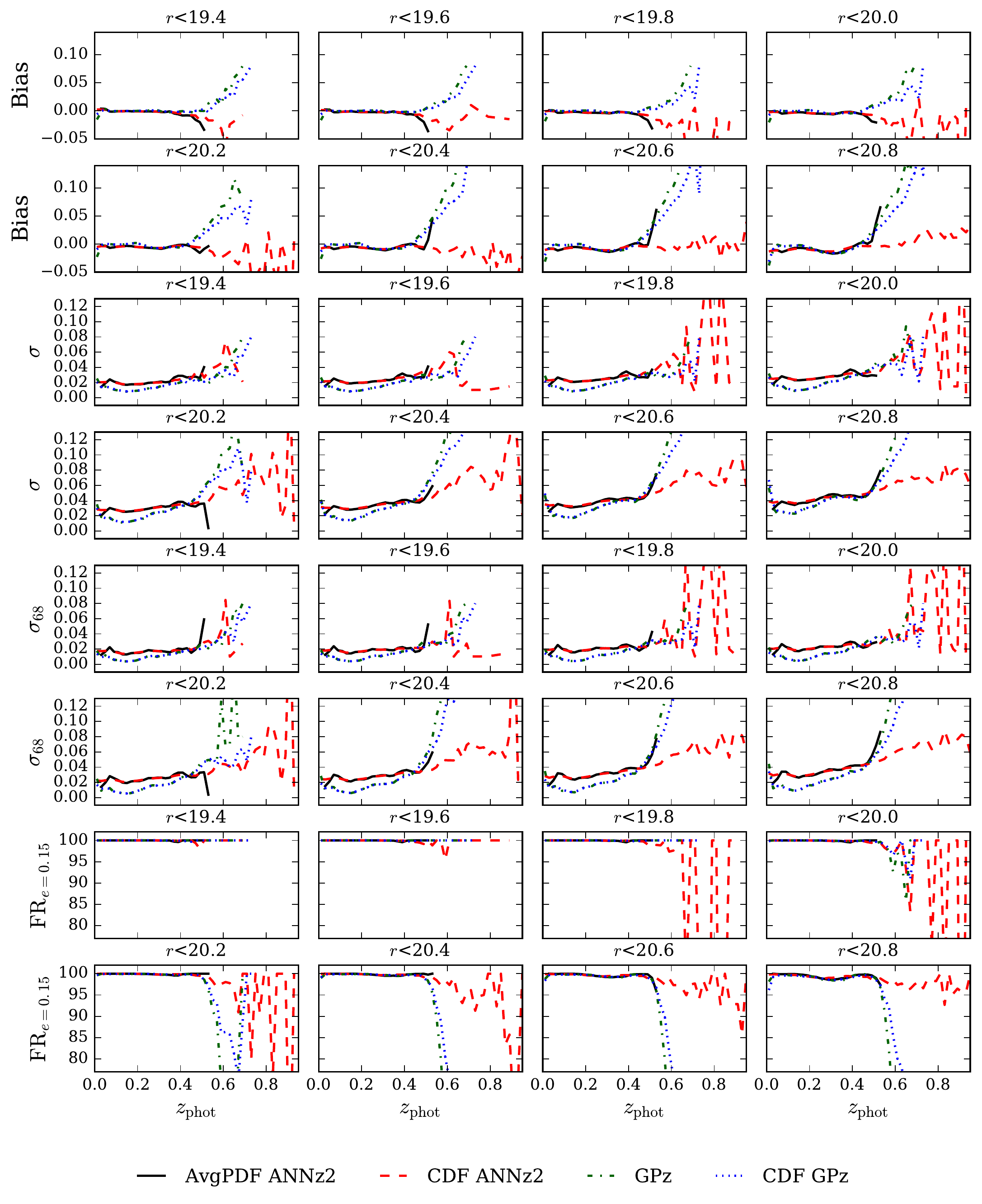}
 \caption{Bias, $\sigma$, $\sigma_{68}$ and $\textrm{FR}_{e=0.15}$ values in the range $0<z_{\rm phot}<1$ for $r\in\left[19.4,20.8\right]$ by using the data of the mock catalogue.
Note that the photometric redshift estimators have good metric values in the range $0\lesssim z_{\rm phot} \lesssim 0.4$ for all $r$-band cuts. 
In this redshift range the metrics slightly worse for deeper cuts.}
 \label{DP:metrics_redshift}
\end{figure*}

In Fig.~\ref{DP:metrics_redshift} we show the bias, $\sigma$, $\sigma_{68}$ and $\textrm{FR}_{e=0.15}$ values in the range $0<z_{\rm phot}<1$ for $r\in\left[19.4,20.8\right]$ by using the data of the mock catalogue.
We see that the photometric redshift estimators have good metric values in the range $0\lesssim z_{\rm phot} \lesssim 0.4$ for all $r$-band cuts. 
In this redshift range the metrics slightly worse for deeper cuts.
On the other hand, we note that the bias and scatter computed for the estimators based on Gaussian GPz grow faster than the CDF-ANNz2 estimator in hight redshift and this is more evident for $r>20.0$.
The above assertion is  in agreement with the discussion performed in Section~\ref{performance_degradation}.

\bsp	
\label{lastpage}
\end{document}